# Unraveling the Roles of Shallow, Deep and Auger Trapping in Charge Carrier Recombination in Triple-Cation Perovskites

*Jitendra Kumar[1], Thomas Kirchartz[5,6], Alexandr Marunchenko[1], Alexander Kiligaridis[1], Shraddha M. Rao[1], Shivam Singh[2,3], Ankur Yadav[4], Monojit Bag[4], Yana Vaynzof[2,3], Ivan G. Scheblykin[1*]*
*Corresponding author(s). E-mail(s): ivan.scheblykin@chemphys.lu.se

[1] Chemical Physics and NanoLund, Lund University, P.O. Box 124, 22100 Lund, Sweden
[2] Leibniz Institute for Solid State and Materials Research Dresden, Helmholtzstraße 20, 01069 Dresden, Germany
[3]Chair for Emerging Electronic Technologies, Technical University of Dresden, Nöthnitzer Str. 61, 01187 Dresden, Germany
[4] Advanced Research in Electrochemical Impedance Spectroscopy Laboratory, Indian Institute of Technology Roorkee, Roorkee, 247667, India
[5] IMD-3 Photovoltaics, Forschungszentrum Jülich, Jülich, Germany
[6] Faculty of Engineering and CENIDE, University of Duisburg-Essen, Duisburg, Germany

## Abstract

Understanding charge-carrier recombination in metal halide perovskites is essential for accurately identifying the factors limiting solar cell efficiency, yet it remains challenging due to the interplay of multiple competing processes. Here, we combine time-resolved photoluminescence and excitation-dependent photoluminescence quantum yield measurements over a wide range of fluences and repetition rates to investigate recombination dynamics in triple-cation perovskite thin films. By jointly analyzing these multidimensional datasets, we develop a unified model that quantitatively reproduces both photoluminescence decays and absolute quantum yields across all excitation conditions. Our results reveal the coexistence of deep and shallow traps, as well as a second-order nonradiative recombination pathway attributed to Auger-assisted trapping. Importantly, this mechanism dominates under one-sun illumination, making it a critical limiting factor for photovoltaic performance. These findings provide a comprehensive framework for understanding recombination in perovskites and highlight the importance of higher-order defect-mediated processes in determining their efficiency.

## 1. Introduction

Defects in semiconductors are local deviations from an ideal periodic crystal structure. These imperfections can trap charge carriers, facilitate non-radiative recombination, reduce carrier mobility, or dope a semiconductor, making it *p*-type or *n*-type. The role and necessity of defects in modulating the conductivity and electrostatic landscape of semiconductors have been central to most modern (opto-) electronic applications.[1] The elimination of such defect states that primarily lead to trapping and recombination of charge carriers has been crucial for high-performance optoelectronic applications—such as solar cells, light-emitting diodes (LEDs), and photodetectors. However, some electronic devices currently under investigation, such as memory cells, memristors, and photonic neuromorphic devices, benefit from the controlled presence of defects because of their functional roles in data storage and signal processing. [2–5].

The presence of traps can be probed using various optoelectronic techniques. [6–11] Among these, bandgap photoluminescence (PL) stands out as the most straightforward and most accessible method for assessing charge trapping and recombination. PL intensity is directly affected by the concentration of free charge carriers, which is altered by trap-mediated processes. Thus, by measuring the PL under identical conditions across different materials and comparing the total photon emission, one can qualitatively assess non-radiative recombination and infer defect-related properties. When supplemented with time-resolved PL (TRPL) decay kinetics, the technique enables the extraction of specific trap parameters. [4,12–17]

To rationalize and explain charge-carrier dynamics in the presence of traps, one must combine detailed experimental data with a physical model that accounts for all relevant processes. The simplest model for recombination via singly-charged (acceptor-like or donor-like) defects is the Shockley-Read-Hall (SRH) model.[18] It considers a single type of trap characterized by four parameters: trap density $N_t$, trap depth $E_t$, electron capture coefficient $k_t$, and hole capture coefficient $k_n$.[18] Notably, the model does not have detrapping (or emission) coefficients as independent parameters, as they follow from the capture coefficients, the trap depth, and the effective densities of states via a detailed balance argument. This assumes that the capture coefficients do not depend on carrier density and are therefore the same in thermodynamic equilibrium and in non-equilibrium situations.

For deep traps, the detrapping rates are sufficiently small to be neglected, simplifying the equations. In contrast, shallow traps are defined by significant detrapping rates to one of the bands. In such cases, detrapping must be explicitly included, as neglecting it would alter the mathematical description and violate detailed balance, and hence thermodynamic consistency. [21,22]

In more realistic scenarios involving both shallow and deep traps, along with radiative and Auger recombination, the total number of parameters can quickly exceed 9. Clearly, extracting so many parameters from limited data, such as PL lifetimes or PL quantum yields (PLQY) alone, leads to significant ambiguity.[17] This ambiguity is a likely reason for the wide variability in reported trap densities and capture coefficients for even nominally identical perovskite materials.[6,10,19,20]

The inherent complexity and nonlinearity of recombination processes therefore require rich, multidimensional experimental datasets to constrain and validate physical models meaningfully. A common strategy is to record TRPL decay curves across a broad range of excitation pulse fluences and fit all data simultaneously.[4,5,17,21] However, such decay curves are often normalized,[12,14,16] discarding information about the absolute PLQY. This loss of information severely limits the ability to identify and quantify key recombination pathways. It may lead to over- or underestimation of critical parameters, or even omission of essential recombination processes when oversimplified models are used.

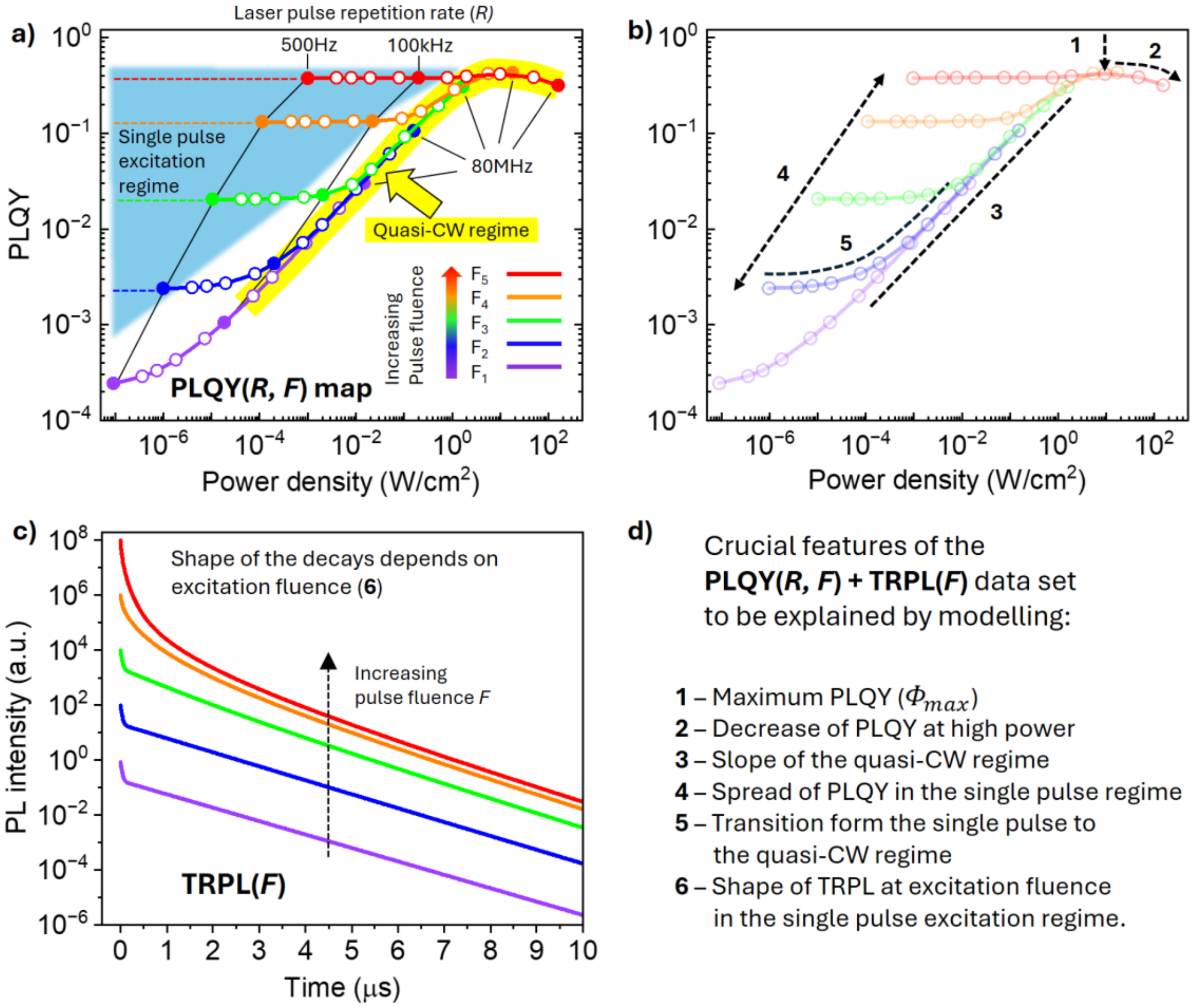

**Figure 1.** Illustration of the complete set of measurements combining PLQY and time-resolved PL data. (a) PLQY(*R, F*) map calculated according to the SRH model previously used for $MAPbI_3$ film.[17] Each color corresponds to a pulse fluence (*F)* according to the color legend, data points connected by a line are measured for different laser pulse repetition rates (*R*, some values are indicated) but the same pulse fluence according to the color legend. X-axis is the time-averaged excitation power density that is proportional to the product *R·F*. The blue shaded region – single pulse excitation regime where PLQY does not depend on *R*, yellow shaded region shows the quasi-CW excitation regime where PLQY depends only on the average power density (*W* ~ *R·F*). (b) Illustration of the critical features **1-5** of the PLQY(*R, F*) map which must be reproduced by theoretical models. c) A sketch of PL decays at different excitation fluences (*F)* which are typically non-exponential. Their shapes (feature **6**) must be fitted together with the PLQY map. d) Description of the features of the PLQY(*R*, *F*) map and TRPL(*F*).

Another common challenge in PL experiments is choosing the appropriate excitation laser repetition rate. This happens because perovskites exhibit slow trap-mediated recombination, and its timescale can overlap with the laser excitation period. For example, a commonly used repetition rate of 100 kHz corresponds to a period of 10 µs, which is shorter than some of the reported trap-mediated timescales in perovskites.[22] This leads to a situation in which the accumulated effect of multiple excitation pulses might govern the system response. Such conditions should be properly modeled to extract meaningful physical parameters.[4,5,17,23–25]

To overcome these limitations, a self-consistent experimental strategy has recently been proposed.[17] As illustrated in Figure 1, it combines multiple TRPL decays measured at varying pulse fluences (*F)* (Figure 1c) with a two-dimensional PLQY map (PLQY(*R*, *F*), Figure 1 a), where PLQY is recorded as a function of both fluence (*F*) and repetition rate of the laser pulses (*R*). By spanning a broad range of excitation conditions, this method generates a highly informative dataset with many features illustrated by numbers in Figure 1b and explained in Figure 1d. These features tightly constrain the parameters of any charge-recombination model that aims to describe the experiment. [17]

It is easier to interpret the two-dimensional PLQY (*R, F*) function when the PLQY (*R, F*) is plotted as a function of average excitation power density (proportional to the product of R and F), with data points corresponding to the same pulse fluence *F* joined by a single curve as shown in Figure 1a. Each line (colored differently) corresponds to a fixed pulse fluence, while each data point represents a different laser pulse repetition rate. This PLQY map (or family of curves) enables the identification of the *single-pulse excitation regime,* where PLQY does not depend on the repetition rate (Figure 1a, blue shaded area). In the single pulse excitation regime, by the time the next excitation pulse arrives, all memory of the previous excitation pulse (*e.g.,* population of trapped charges) decays completely.[4,17] The other extreme is the quasi-CW excitation regime (Figure 1a, yellow shaded regime). In this excitation regime, the PLQY is the same as one would get under a continuous wave (CW) excitation of the same average power density. It occurs when the repetition rate is so high that the population of long-living species (e.g., trapped charges) is close to steady state. Visually, all curves for each specific *F* merge at sufficiently high *R* into a single curve, which corresponds to the quasi-CW regime. In this regime, PLQY depends only on the time-averaged power density.[17] Figure 1a shows an example of how PLQY(*R, F*) might look; it was calculated using the SRH+ model and the previously used parameters for $MAPbI_3$ perovskite.[17] Depending on the nature of the recombination pathways in a semiconductor, the shape of the PLQY map as well as the shape of the TRPL curves can be very different, serving as fingerprints of the charge carrier dynamic pathways in the particular material.[4,17,21]

In this work, we apply this advanced experimental and modelling approach to thin films of the triple-cation perovskite $Cs_{0.05}(FA_{5/6}MA_{1/6})_{0.95}Pb(I_{0.9}Br_{0.1})_3$, (hereafter referred to as TC). We demonstrate that the PL dynamics in TC cannot be adequately described within the conventional SRH framework assuming a single trap type, nor by models incorporating only shallow and deep traps. A consistent description of both the PLQY(*R, F*) map over an exceptionally broad range of average excitation power densities (0.0002–1600 Suns) and the TRPL kinetics measured across three orders of magnitude in pulse fluence is achieved only when a bi-molecular non-radiative process—specifically, Auger trapping—is also included in the model.

The role of each recombination pathway is analyzed in detail using an in silico passivation approach over a wide range of charge-carrier concentrations. We show that the bi-molecular non-radiative processes, such as Auger trapping, are likely to represent a fundamental limiting factor for the solar cell efficiency of TC-based devices.

## 2. Experiments and simulations

We used thin TC perovskite films with a thickness of only 50 nm (see supporting information (SI) section 1.1 for the preparation procedure), which is 4-6 times smaller than typically used for solar cells. Such a small thickness implies that the sample had a low optical density (Figure S2b), thereby minimizing contributions from charge diffusion to the PL dynamics, as explained in detail below. In contrast, for typical perovskite photovoltaic films used for making solar cells (≈ 400 – 600 nm), diffusion cannot be neglected,[26,27] and should generally be considered in the interpretation of the full TRPL and PLQY(*R, F*) data set. This additional factor makes theoretical modelling complex and highly ambiguous because charge trapping and diffusion can produce similar experimental signatures. Choosing such thin films allowed us to avoid this uncertainty.

To perform PLQY (*R*, *F*) mapping and acquire excitation fluence-dependent PL decay kinetics, a custom-built setup based on an inverted wide-field PL microscope was used (SI section 2).[17] The sample was excited by a 482 nm diode laser with a 200 ps pulse width through a 40× (NA = 0.6) objective lens. The PL signal was collected using the same objective lens and cleaned of excitation light by a set of interference filters. PL was detected either with an EM-CCD to measure PLQY(*R, F*) maps (see SI section 3) or with a hybrid photomultiplier detector. The latter detector, connected to a time-correlated single-photon counting (TCSPC) module, enabled the measurement of TRPL decay kinetics. The instrumental response function of this setup was approximately 200 ps. The setup was calibrated to measure the absolute PLQY as described in detail in Section 4 of the SI.

Using this setup, the PLQY(*R, F*) map was measured using pulse fluences $F_1 = 3.4 \times 10^8$, $F_2 = 4.0 \times 10^9$, $F_3 = 4.3 \times 10^{10}$, $F_4 = 5.4 \times 10^{11}$, and $F_5 = 4.9 \times 10^{12}$ photons/cm² while the laser pulse repetition rate *R* was scanned from 1 kHz to 80 MHz. TRPL decays were measured for four laser pulse fluences ($F_2$, $F_3$, $F_4$, and $F_5$) at a 10 kHz pulse repetition rate (Figure 2). See SI section 3 for all experimental parameters. The charge-carrier concentrations generated by pulses of different fluences are listed in Table S1. The experimental data were modeled using various charge-recombination models implemented in a home-written MATLAB program. The program calculates the fluence-dependent decays and PLQY(*R, F*) maps by solving a set of coupled differential equations by a fourth-order Runge–Kutta method. The solid lines in Figure 2 represent the best-fit of the experimental data by a model described in detail below.

## 3. Results and discussion

### 3.1 The full model explaining TRPL decays and PLQY(*R, F*) of the TC film at the same time

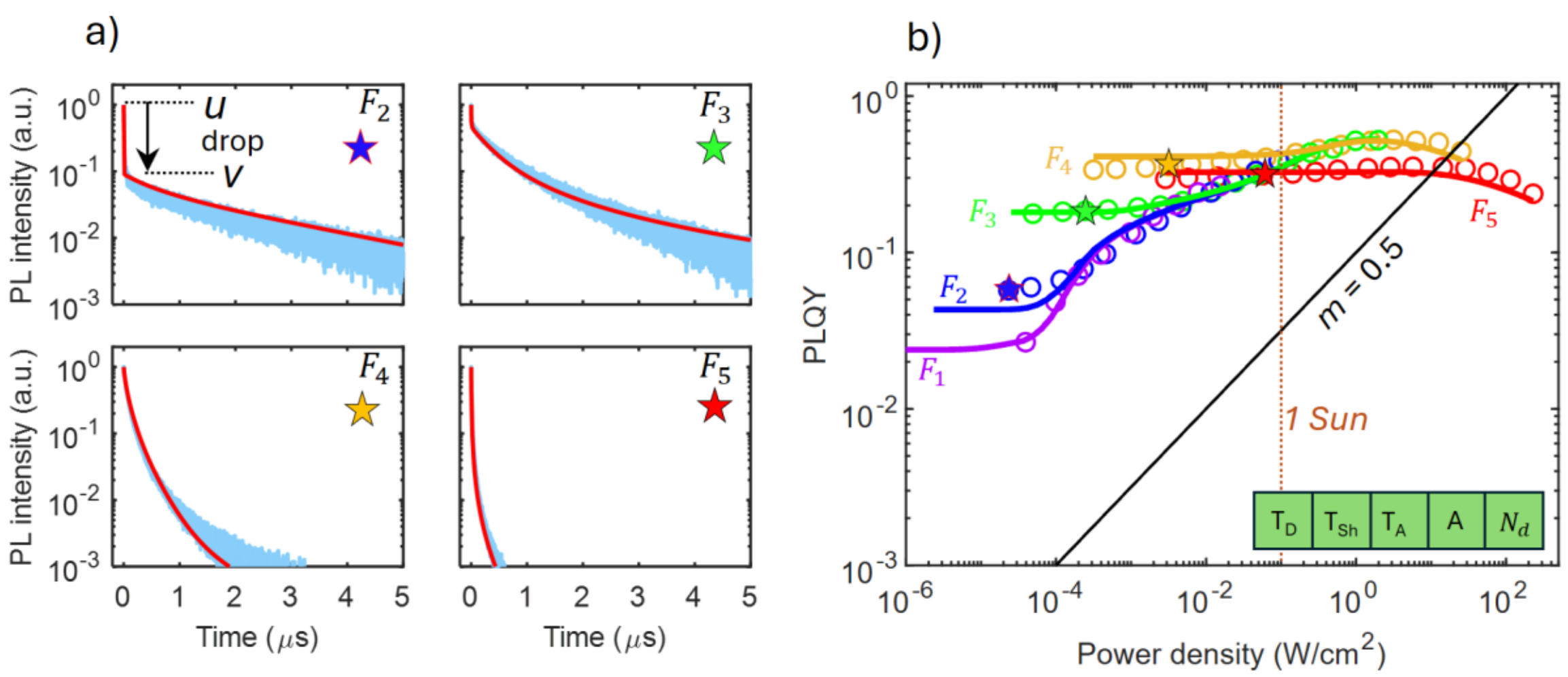

**Figure 2.** Experimental and modelled data for 50 nm TC perovskite film. a) The experimental TRPL decay (blue) along with the fits (red), labels $F_2$, $F_3$, $F_4$ and $F_5$ show the pulse fluence used for each TRPL measurement at 10 kHz repetition rate. b) PLQY(*R, F*), where PL quantum yield $\Phi$ is plotted against average power density $W$ for different values of $R$ and $F$, circles - experimental data with the color designating the pulse fluence ($F_1$– $F_5$, see the labels), solid lines - calculated $\Phi$ (the same color code of the pulse fluence). The stars of each color show the measurement conditions and $\Phi$ corresponding the TRPL decays plotted in (a). The labels in the low right corner of (b) show the active processes besides radiative recombination considered in the model (see the text for details), the solid line labeled with $m$ = 0.5 is the reference line showing the power law $W^{0.5}$.

It was found that, to explain both PL decays and PLQY(*R, F*) maps within a single model using the same parameters, all the processes illustrated in Figure 3 should be considered. Below, we describe the model, referred to hereafter as the full model, and motivate the necessity of each included process. The model (see SI section 9) contains three types of acceptor-like charge trapping defects ($T_D$, $T_{Sh}$ and $T_A$), all traps are electron traps in this model. $T_D$ is a deep trap with strong asymmetric capture coefficients (electron capture is much faster than hole capture). The coefficients are such that the defects $T_D$ can saturate and become filled with electrons within the range of excitation fluences used experimentally. $T_{Sh}$ is an acceptor-like shallow trap located close to the conduction band, which (in contrast to deep traps) allows the de-trapping of electrons back into the conduction band. Both $T_D$ and $T_{Sh}$ cause *p*-photodoping.[13] $T_A$ is a so-called Auger trap, which can trap an electron (*n*) only in the presence of a hole (*p*) nearby. This makes the trapping process (called Auger trapping) second-order in concentration.[17,28–30] In addition to these three types of trapping, the model also considers second-order radiative recombination, third-order Auger recombination (*n* + *p* + *p*), and *p*-type doping (*n* < *p*). Although the model uses 13 parameters (see SI Table S3), we will explain below why this complexity is necessary to explain all features of the dataset (Figure 2).

### 3.2 Shallow trapping as the origin of the initial drop in TRPL decay

TC films exhibit two distinct components in their TRPL decays measured at low pulse fluences ($F_2$): i) the fast component that lasts up to a few tens of nanoseconds appearing as the initial drop of PL intensity from level *u* to level *v* with relative amplitude $\xi = (u - v)/u$, see Figure 2a, and ii) a much slower decay at a time scale of a microsecond that follows the initial intensity drop. This behavior is the principal difference from $MAPI_3$ film, where monoexponential PL decays are often observed at low fluences.[17,31–33] The presence of two components in the PL decay of TC does not allow for explaining the results with the same model (deep trap, Auger recombination, and Auger trapping) that was used to describe the same experiments for $MAPbI_3$ previously[17], as illustrated in SI Section 10.1.

In the literature, one can find three common explanations of a fast initial PL decay component that is followed by slower decay components at later times:[23,34–37] i) fast initial trapping by deep traps which get saturated, i.e. only slowly capture the second carrier thereby leading to a slow longer time component, ii) charge carrier diffusion and homogenization of their distribution over the film thickness, and iii) fast charge carrier trapping by shallow traps, where again the capture of the second carrier is slow. Below, we will consider all three cases in the context of our experiments.

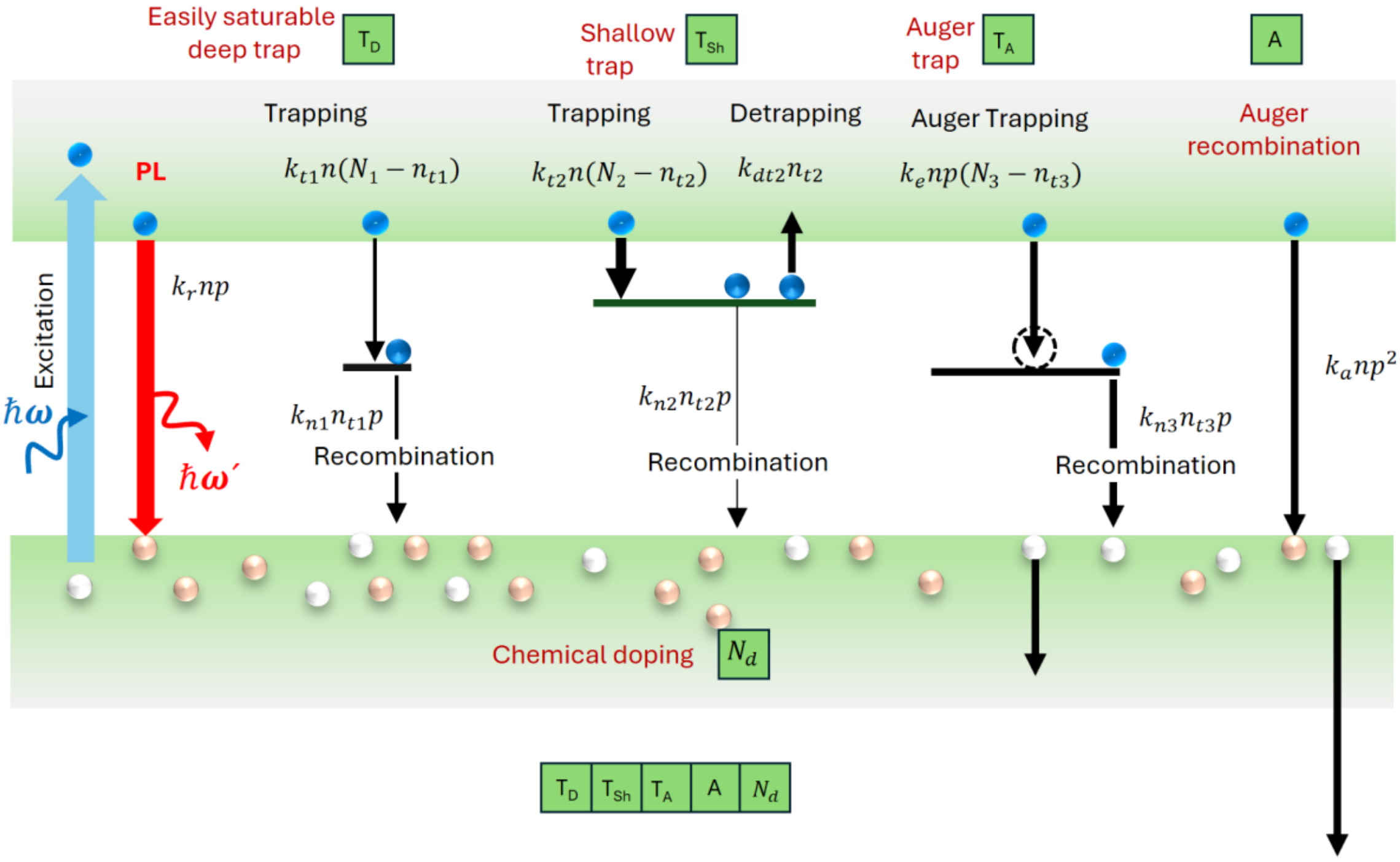

**Figure 3**. The full model required to fit experimental data for the TC film. Besides radiative recombination it considers an easily saturable deep trap ($T_D$), a shallow trap ($T_{Sh}$), a deep trap which captures a free electron by Auger trapping ($T_A$), radiative recombination, Auger recombination (A), and chemical *p*-doping ($N_d$). These processes are shown as pictograms in the bottom of the figure and these pictograms will be used in the other figures in the paper. The designations of the model parameters can be found in SI.

An easily saturable deep trap can, in principle, produce a highly non-exponential PL decay at low fluence in the single pulse excitation regime. This happens because the trapping rate is proportional to the density of empty traps and therefore it decreases as the traps become filled. Once saturation is reached, electron trapping slows down, resulting in a long-lived PL component at later times.

It is possible to fully reproduce the TRPL decay at fluence *F2* including the initial drop with ξ≈0.9 using the saturable deep-trap model, setting very large (>$10^5$ times) difference between the electron ($k_t$) and hole ($k_n$) capture coefficient. Then, the initial PL intensity drop reflects the fraction of electrons captured needed to saturate the traps. In this case, increasing the fluence by 10 times (from *F2 to F3*) should reduce the initial drop by approximately the same factor of 10 leading to ξ≈0.09 (Figure S11b).However, at *F3* experimentally the PL intensity initially drops to the level of 0.55 (ξ≈0.45), which means about 5 times larger drop as predicted.

If instead the TRPL decay at *F3* is fitted (Figure S11d), reducing the fluence by a factor of 10 (from *F3 to F2*) should move the system out of the saturation regime, and the slow decay component should no longer be present. This is inconsistent with the experimental observation that the slow component persists at fluence F2 (Figure S11c). These inconsistencies indicate that the initial PL drop observed at *F2* and *F3* cannot be solely explained by saturation of deep traps.

The initial fast component could, in principle, also arise from diffusion-induced homogenization of charge carriers across the film thickness shortly after photogeneration (SI section 5). Owing to the high absorption coefficient of metal halide perovskites (absorption length <50 nm for blue light),[38,39] excitation at 482 nm produces a strongly inhomogeneous carrier distribution in thick films (300 nm) typically studied in the literature. If diffusion through the film is slower than the TRPL experiment's time resolution, a drop in PL intensity should occur on the diffusion timescale, as PL intensity depends directly on carrier concentration.

To minimize this effect, we used TC films only 50 nm thick, with an optical density of ~0.35 at our excitation wavelength. In this case, the initial carrier concentration varies by only about a factor of two across the film thickness, and diffusion alone can account for at most ~10% reduction in PL intensity (Figure S10, SI section 5). Experimentally, however, we observe an initial drop to the level of 0.1 corresponding to ξ≈0.09 (Figure 2a), indicating that diffusion cannot account for this rapid initial PL decay.

We are then left with the shallow-trapping mechanism. Indeed, in the presence of only shallow traps ($T_{Sh}$), two distinct components should be observed in the PL decay of any semiconductor.[37] The fast component arises from fast trapping of the charge carriers (electrons, as assumed here). The fast decay ends when an equilibrium is established between the trapping and de-trapping processes. After that, the equilibrated population of electrons and trapped electrons decay slowly via non-radiative recombination of trapped electrons with free holes and radiative recombination of the free electrons and holes. Both processes are very slow at low excitation conditions, resulting in a slowly decaying tail in TRPL, that quantitatively agrees with our data. Therefore, for our samples, shallow trapping is likely the primary reason for the peculiar shape of the TRPL decay at low excitation fluence.[23,24]

### 3.3 Shallow traps alone cannot explain TRPL together with PLQY

The presence of shallow traps ($T_{Sh}$) can explain the initial rapid drop in PL intensity and the general presence of both fast and slow components in PL decay under low fluence. Moreover, if we ignore PLQY, we can nicely reproduce the complete PL decay at low fluence ($F_2$). However, such a simple model (model SRH-2, SI section 10.2) fails to account for the PL decay shapes at higher fluences and the PLQY(*R, F*) map, as illustrated in Figure S15. In this model, the slow component of the PL decay at low fluence is primarily defined by the nonradiative recombination rate of trapped electrons at the trap level and free holes in the valence band. To reproduce the experimental decay rate of the slow component, we needed to set a relatively high $k_n$ from the shallow trap level, resulting in PLQY much lower than experimentally observed at all excitation conditions (Figure S15). Apart from this, with increasing pulse fluence, the PL decay became too fast (owing to their non-linear recombination kinetics) and incompatible with the experiment (see details in SI section 10.2). Thus, the charge recombination in our TC samples cannot be attributed solely to shallow traps.

### 3.4 The need for deep traps

To overcome these difficulties in fitting the experimental data, additional processes should be incorporated into the model. The obvious thing to try is to include a deep trap that introduces first-order non-radiative recombination. Presence of deep traps is widely discussed in literature on MHP, and an SRH model with only deep traps has been rather successfully used to explain the same experiment as discussed here (TRPL and PLQY(*R, F*)) for a simpler MHP system – a film of $MAPbI_3$.[17]

Our data also provides direct evidence for deep traps. At low fluence ($F_2$), the PL decay shows an exponential tail, as seen in the constant differential decay time between 1–8 µs (Figure S12). Such behavior cannot arise from shallow traps alone, which would produce a power-law tail,[23,24] indicating the presence of at least one first-order process. A natural explanation is recombination via deep traps, where detrapping is negligible due to the large energy gap ($\Delta E \gg kT$) between the nearest band edge (conduction or valence band) and the trap level. PL lifetime together with the observed PLQY also suggests that the capture rate of the deep traps, as expected, is slower than the capture rate of the shallow ones,[29] making it consistent with the hundreds-of-nanoseconds timescale of the slow PL component at low fluence. [17]

### 3.5 Signature of Auger recombination processes at high excitation powers

The observed decrease of PLQY with increasing excitation power at high excitation densities (feature 2 in Figure 1) requires the presence of third- or higher-order non-radiative recombination processes. Such processes compete with radiative recombination (which is second order in carrier concentration) and become increasingly dominant at elevated excitation levels.

The standard approach in literature is to invoke Auger recombination, a third-order process.[17,28,30] In our case, a suitable Auger process involves the recombination of a free electron with a free hole in the presence of a second hole. The latter receives the excess energy released during the recombination (Figure 3). However, including the conventional Auger recombination alone helps quantitatively reproduce the experimental data. As shown in SI section 10.3, this model fails to simultaneously explain the PLQY(*F*,*R*) map and the fluence-dependent TRPL kinetics. Increasing the Auger contribution to reduce the maximum PLQY ($\Phi_{\mathrm{max}}$, feature 1 in Figure 1) to the experimentally observed level leads to an excessive reduction of PLQY at high excitation powers (incompatible with feature 2) and produces TRPL decays at the highest fluence ($F_5$) that are significantly faster than observed. Thus, even with conventional Auger recombination, we still cannot explain the experimental data.

We therefore considered inclusion of a second-order non-radiative recombination pathway[40–44]. Such a process can directly compete with radiative recombination, as both are second order in carrier concentration, and consequently reduce the maximum achievable PLQY. The necessity of introducing this additional second-order recombination pathway depends critically on the reliability of the absolute PLQY values. Accurate determination of absolute PLQY is inherently challenging due to multiple systematic and random errors, which we quantify in SI section 4. We estimate an uncertainty of ±30% in absolute PLQY, meaning that the entire PLQY map may shift vertically by ±30%. In contrast, the relative accuracy—i.e., the shape of the PLQY map—is significantly higher (±10%) and is mainly limited by sample instability. [21] Even within this uncertainty, the maximum PLQY for our TC sample ($\Phi_{max}$, feature 1 in Figure 1) lies in the range 0.35–0.65, well below unity, thereby justifying inclusion of an additional bimolecular non-radiative process.

A suitable mechanism of bimolecular non-radiative recombination is Auger trapping – a process recognized in many semiconductors and semiconductor devices since the 1950s.[17,28,30,42–45] In this process, charge trapping is assisted by the presence of another free carrier. While historically such "non-linear" trapping has been considered secondary to the usual "linear" trapping by multiphonon emission in narrow-gap materials like Ge and Si,[44] recent first-principles modeling and experimental signatures in GaN, InN, and β-$Ga_2O_3$ demonstrate that Auger trapping can dominate losses in these wide-bandgap systems.[41–43,46] Several experimental studies on metal halide perovskites have also reported signatures of second-order non-radiative recombination, and this topic has been receiving increasing attention in recent years.[47–49]

In our model, a feasible mechanism is when a free electron interacts with a free hole in the presence of an empty electron trap, resulting in electron trapping while the hole gains excess energy. Below the trap-saturation regime, the concentration of empty traps remains approximately constant, rendering this mechanism effectively second-order with respect to free-carrier concentrations. As a result, Auger trapping scales with excitation power in the same way as radiative recombination and competes directly with it, thereby limiting $\Phi_{\max}$ to the experimentally observed level. Importantly, Auger trapping does not significantly affect TRPL decay at the highest fluences because it is slower than the conventional Auger recombination and requires empty traps, which become progressively saturated at high carrier densities. Including Auger trapping in the model substantially improved overall agreement with the experimental data (SI Section 10.4); however, several important discrepancies remained.

### 3.6 Signature of chemical doping in the sample leading to increased PLQY at low pulse fluence.

Another factor affecting PL dynamics that is commonly discussed but has not been considered yet in our model is chemical doping. It is generally accepted that unintentional doping levels in MHPs are relatively low;[50,51] however, at low excitation power regimes, doping at the level of $10^{15}$ $cm^{-3}$ can make a significant difference.

In the presence of chemical doping, the PL intensity in the single pulse excitation regime $I_{PL} \propto \Delta n_0(\Delta n_0 + N_d)$, where $\Delta n_0$ is the initial concentration of free charge carriers (electrons or holes) created by the laser pulse, (section S7) and $N_d$ is the concentration of holes always present in the system (we consider *p*-doping here). We refer to the latter as chemical doping to distinguish it from photodoping, which arises from asymmetrical trapping of charge carriers. According to this equation, an increase in $N_d$ should increase the radiative decay rate and, potentially, the PLQY. On the other hand, $N_d$ also increases the non-radiative recombination rate of trapped electrons with free holes, which, in the case of shallow traps, can decrease PLQY and cause the long tail (later part of the decay) of TRPL decay faster at low fluences ($F_2$, $F_3$). Therefore, the influence of doping is expected to be complex and dependent on the excitation regime, which may help to explain our data. We also obtained experimental evidence of chemical doping from a multi-pulse TRPL experiment (see SI section 7, Figure S13).[4,22] By adding chemical doping, we completed the full model (Figure 3) and were able to fit the whole data set consistently (Figure 2).

### 4.1 *In silico* passivation of different recombination processes.

To further rationalize and illustrate the impact of different processes on the TRPL and PLQY, an ideal approach would be to remove a selected process and observe its effect on the experimental data. However, eliminating individual charge-recombination mechanisms from a film in a controlled manner is practically impossible. Here, simulations are beneficial as they allow us to remove any desired mechanism from the complete model that explains the data. This procedure can be seen as *in silico* passivation; the results are presented in Figure 4.

Figure 4(a,b) shows the effect of removing the easily saturable deep traps ($T_D$) from the model. One can see that these traps affect the TRPL only at low fluences ($F_3$, $F_2$, and lower), making the slow component

of the PL decay faster and decreasing PLQY in the low average excitation power density regime when they are not yet saturated (power density < $10^{-2}$ W/cm$^2$ = 0.1 Sun). The ease of their saturation makes these deep traps benign for solar cells, since they do not influence charge recombination at average excitation powers of 0.1 - 1 Sun.

Figure 4 (c,d) shows the effect of passivating the shallow traps ($T_{Sh}$). One can see that shallow traps make the PL decay drastically non-exponential by inducing a rapid decay component (the initial PL intensity drop) in the TRPL at low fluences ($F_2$, $F_3$) and, at the same time, creating a very slow decaying tail (at the microsecond time scale). Thus, the very long lifetime components often observed in perovskites are most likely due to the presence of shallow traps, rather than their absence (because removing shallow traps makes the TRPL decay faster). Shallow trapping also decreases PLQY, but only at power densities below $10^{-2}$ W/cm$^2$=0.1 Sun. Thus, in terms of reducing PLQY, shallow trapping is expected to have a minor effect on the solar cell performance. Note, however, that shallow trapping increases the overall charge-carrier lifetime.

Upon removing the *p*-doping from the full model (Figures 4e,f), we observe that *p*-doping makes the tail of TRPL at low fluences ($F_2$, $F_3$) decay faster. As for PLQY, chemical doping at this level ($N_d \approx 3\times10^{15}$ cm$^3$) is essential only for the lowest fluences ($F_1$, $F_2$ and $F_3$) and relatively low pulse repetition rates because only in these conditions $N_d$ is larger or comparable to the concentration of the photogenerated carriers (see Table S1). It is interesting to note that, within the framework of our model, *p*-type doping decreases PLQY. This happens because doping increases both radiative and non-radiative rates, and its effect on PLQY depends on which effect is stronger. As for the impact on photovoltaic applications, PL quenching from chemical doping at this level is negligible, as it has little effect on PLQY under solar cell operating conditions.

Figure 4(g,h) shows the effect of passivation of the Auger recombination (A). Auger recombination reduces the PLQY for all fluences ($F_1$–$F_5$) once the average power density exceeds 0.1 W/cm$^2$ (>1 Sun). At higher fluences ($F_4$ and $F_5$), however, it leads to a reduction in PLQY under all conditions (all repetition rates). TRPL is noticeably affected at the highest fluence $F_5$ only. Therefore, this process alone is not particularly significant under solar cell operation conditions.

In contrast to regular Auger recombination, Auger trapping is very important for PLQY over a broad range of regimes. Figure 4(i,j) shows the effect of passivation of the Auger trap (trap $T_A$). One can see that the effect of Auger trapping on the PLQY(*R, F*) map is substantial over a wide range of powers from $10^{-4}$ to 1 W/cm$^2$ (from 0.001 to 10 Suns) which makes this process highly relevant for photovoltaics. In terms of pulse fluence, it is remarkable that Auger trapping is important even for the lowest fluence $F_1$ as soon as the repetition rate exceeds 1 MHz. Note also that Auger trapping affects TRPL at all measured pulse fluences. However, at very high fluences and high average charge-carrier concentrations, the usual Auger process (third order) dominates over Auger trapping (second order, and even less due to trap saturation). Consequently, PLQY at the highest fluence ($F_5$), as well as at $F_4$ (as soon as the repetition rate exceeds 10 MHz) is unaffected by Auger trapping.

Now that we understand the importance of the Auger processes, let's remove everything from the model except for radiative recombination, Auger recombination, and Auger trapping (Figure 4 (k,l)). It appears that even this truncated model can explain all experimental data obtained for the average power densities above 0.5 Sun (marked by green boxes in the same Figure). This means that at full solar illumination, only non-linear recombination processes are important, because all the usually considered defect-related processes are either saturated or cannot compete with higher-order processes. However, since the crucial process here is Auger trapping, the presence of defect states remains critical, although the trapping mechanism differs. Thus, our results show that second-order non-radiative recombination (e.g., Auger trapping) is an important factor limiting the efficiency of triple-cation perovskite solar cells.

Note that the only difference between the TC films used for solar cells[52–54] and the film studied here is the lower thickness of the latter (50 nm), which was chosen to exclude the contribution of charge diffusion to PLQY and TRPL kinetics and make the presented analysis possible. Although these results may not quantitatively represent all triple cation samples processed under different conditions, we believe they clearly demonstrate the need to consider second-order non-radiative processes, such as Auger trapping, seriously in the context of solar cell performance and the theoretical description of charge dynamics in MHP in general. As for the effect of the thickness alone, in typical 300 nm thick perovskite films used in solar cells short absorption length and diffusion will lead to several times lower equilibrated concentrations of charge carriers than in our experiments under the same illumination conditions. However, this rather minor difference will not change our conclusion on the importance of Auger trapping processes.

At the moment, it is not possible to determine the chemical nature of the defects that assist in Auger trapping in TC perovskite. By looking at the concentration of these types of traps obtained from the modelling, we, however, can conclude that these traps must be common in the material since their concentrations are of the order of $10^{16}$ $cm^3$, being even higher than the concentrations of shallow traps ($\approx 10^{15}$ $cm^3$) and deep traps ($\approx 10^{14}$ $cm^3$) derived from the simulations, see Table S3.

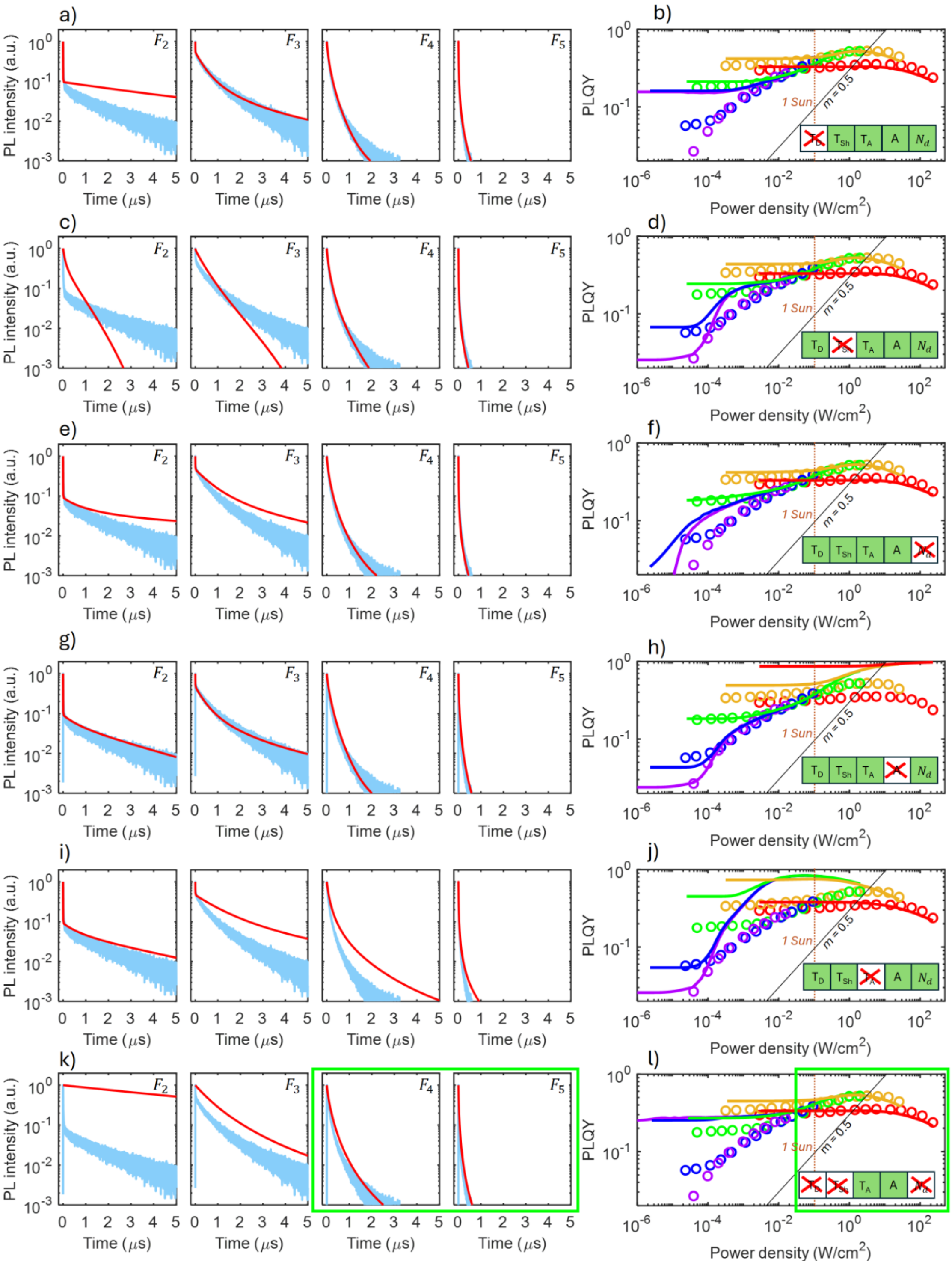

**Figure 4.** The experimental TRPL decays (left) and the PLQY (*R, F*) map (right) plotted together with the corresponding calculations using the full model (which fits the whole data set, see Figure 2), where certain processes were removed (*in silico* passivation): a) and b) - deep traps ($T_D$) are removed, c) and d) the shallow traps ($T_{Sh}$) are removed, e) and f) chemical doping is removed g) and h) Auger recombination is removed, i) and j) – Auger trapping is removed, k) and l) - deep traps, shallow traps and chemical doping are removed leaving only radiative recombination, and Auger recombination and Auger trapping in the model.

## Conclusions

In this work, we conducted a comprehensive experimental and theoretical investigation of charge-carrier recombination in thin films of triple-cation perovskites using time-resolved photoluminescence (TRPL) and fluence- and repetition-rate-dependent photoluminescence quantum yield (PLQY) measurements. By combining these extensive datasets with a robust modelling framework, we developed, for the first time, a unified model that accurately captures both the PL decay dynamics and the absolute PLQY across a broad range of excitation fluences and laser repetition rates. This model not only achieves excellent quantitative agreement with experimental observations but also provides new mechanistic insights into the recombination processes governing these complex materials.
Our analysis reveals the coexistence of three distinct types of traps: i) Deep traps, which are easily saturated and dominate recombination at very low excitation fluences; ii) Shallow traps, which strongly influence TRPL decays at low fluence by introducing both fast and slow components; and iii) Auger traps, which give rise to a second-order non-radiative recombination pathway and become the primary loss channel under solar-relevant excitation conditions.

Auger trapping is identified as the dominant mechanism limiting the PLQY under solar illumination, thus constraining the performance of triple-cation perovskite solar cells. While conventional interpretations often downplay the role of defects under operating conditions—since usual trapping mechanisms are largely saturated at typical solar fluences (as we also showed in our study)—considering bimolecular non-radiative recombination mediated by defects through Auger trapping fundamentally changes this perspective. This mechanism brings defects back to the forefront as a critical factor limiting solar cell efficiency.

Equally important, this work highlights the value of advanced experimental methodologies that generate multidimensional datasets, enabling unambiguous theoretical interpretation of charge-carrier dynamics and guiding future strategies for material and device optimization.

## Author contributions:

JK carried out all experiments, wrote the code for modeling, analyzed the data, calibrated the setup, and drafted the manuscript. AM contributed to data interpretation and PLQY calibration of the setup. IS conceived the idea, interpreted the data together with JK, assisted JK in theoretical modeling, developed the PLQY calibration procedure, adjusted the setup, and wrote the paper in conjunction with JK. AK developed the automation for the PLQY mapping setup. SR contributed to the PLQY mapping protocols and initial measurements. SS prepared the TC samples. YV contributed to the idea of the paper, data interpretation, supervision, and writing of the paper. TK contributed to data interpretation, modelling, and writing. MB and AY contributed to data interpretation and initial experiments with other perovskites, which helped us to understand the TC sample. IS supervised the whole project. All authors edited the manuscript.

## Acknowledgement

The work was supported by the Swedish Research Council (project 2020-03530), via a joint Indo-Swedish project (Swedish Research Council 2018-0764, DST/INT/SWD/VR/P-13/2019), Crafoord Foundation (project 20230552), and NanoLund (project 12-2023). JK thanks the Wenner-Gren Foundation for the postdoctoral scholarship (UPD2022-0132). AM acknowledges Light and Materials profile area at Lund University (Young Investigator Synergy Award, 2024).

**Supporting Information Available:**
Details of the sample preparation, optical setup, and its calibration, PLQY (*R, F*) measurements, the full model, and some simpler models.

**Supporting information to**

# Unraveling the Roles of Shallow, Deep and Auger Trapping in Charge Carrier Recombination in Triple-Cation Perovskites

*Jitendra Kumar[1], Thomas Kirchartz[5,6], Alexandr Marunchenko[1], Alexander Kiligaridis[1], Shraddha M. Rao[1], Shivam Singh[2,3], Ankur Yadav[4], Monojit Bag[4], Yana Vaynzof[2,3], Ivan G. Scheblykin[1*]*
*Corresponding author(s). E-mail(s): ivan.scheblykin@chemphys.lu.se

[1] Chemical Physics and NanoLund, Lund University, P.O. Box 124, 22100 Lund, Sweden
[2] Leibniz Institute for Solid State and Materials Research Dresden, Helmholtzstraße 20, 01069 Dresden, Germany
[3]Chair for Emerging Electronic Technologies, Technical University of Dresden, Nöthnitzer Str. 61, 01187 Dresden, Germany
[4] Advanced Research in Electrochemical Impedance Spectroscopy Laboratory, Indian Institute of Technology Roorkee, Roorkee, 247667, India
[5] IMD-3 Photovoltaics, Forschungszentrum Jülich, Jülich, Germany
[6] Faculty of Engineering and CENIDE, University of Duisburg-Essen, Duisburg, Germany

**Table of Contents**

## 1.Sample preparation

### 1.1 Preparation of triple cation perovskite films

**Materials**

Formamidinium iodide (FAI, $HC(NH_2)_2$), and methylammonium iodide (MAI, $CH_3NH_3I$) were purchased from GreatCell Solar Materials. Cesium iodide (CsI), lead iodide ($PbI_2$) and lead bromide ($PbBr_2$) were purchased from Tokyo Chemical Industry Co., Ltd. (TCI). Anhydrous solvents such as Dimethylformamide (DMF) and Dimethyl Sulfoxide (DMSO) were purchased from Acros Organics.

**Solution Preparation**

To prepare the triple cation films, the precursor solution was prepared by dissolving $PbI_2$ and $PbBr_2$ in a solvent mixture (DMF/DMSO = 4/1) and CsI in DMSO. The $PbI_2$, $PbBr_2$ and CsI solution was heated at 150 °C for 10 minutes. After cooling down, CsI solution, $PbI_2$ and $PbBr_2$ solutions were mixed in a volume ratio of 0.05:0.85:0.15, to obtain an inorganic stock solution. FAI and MAI powders were added and weighed in two separate vials, into which the appropriate amount (0.95:1 molar ratio) of inorganic stock was added. This creates two new solutions, of the formula $Cs_{0.05}(FA\ or\ MA)_{0.95}Pb(I_{0.9}Br_{0.1})_3$. Finally, these two solutions were mixed in a 5:1 v/v ratio, in order to achieve the final molecular formula $Cs_{0.05}(FA_{0.83}MA_{0.17})_{0.95}Pb(I_{0.9}Br_{0.1})_3$. To fabricate a 50 nm thin perovskite film, the prepared precursor was diluted with a mixed DMF:DMSO (4:1) solvent in a volumetric ratio of 1:3.

**Film Fabrication**

The glass substrates were cleaned with deionized water, acetone, and isopropanol by ultrasonication for 10 min in each solvent. The substrates were then dried with $N_2$ and treated with oxygen plasma at 100 mW for 10 min. The substrates were immediately transferred to drybox (relative humidity < 1%) for perovskite deposition. The perovskite layer was deposited via a two-step spin-coating procedure with 1000 rpm for 10 s and 6000 rpm for 30 s. Anhydrous chlorobenzene (150 µL) was dripped on the spinning substrate during the last 5 s of the second spin-coating step. Subsequently, the spin-coated samples were annealed at 100 °C for 30 min.

### 1.2 Preparation of the reference dye films for PLQY calibration.

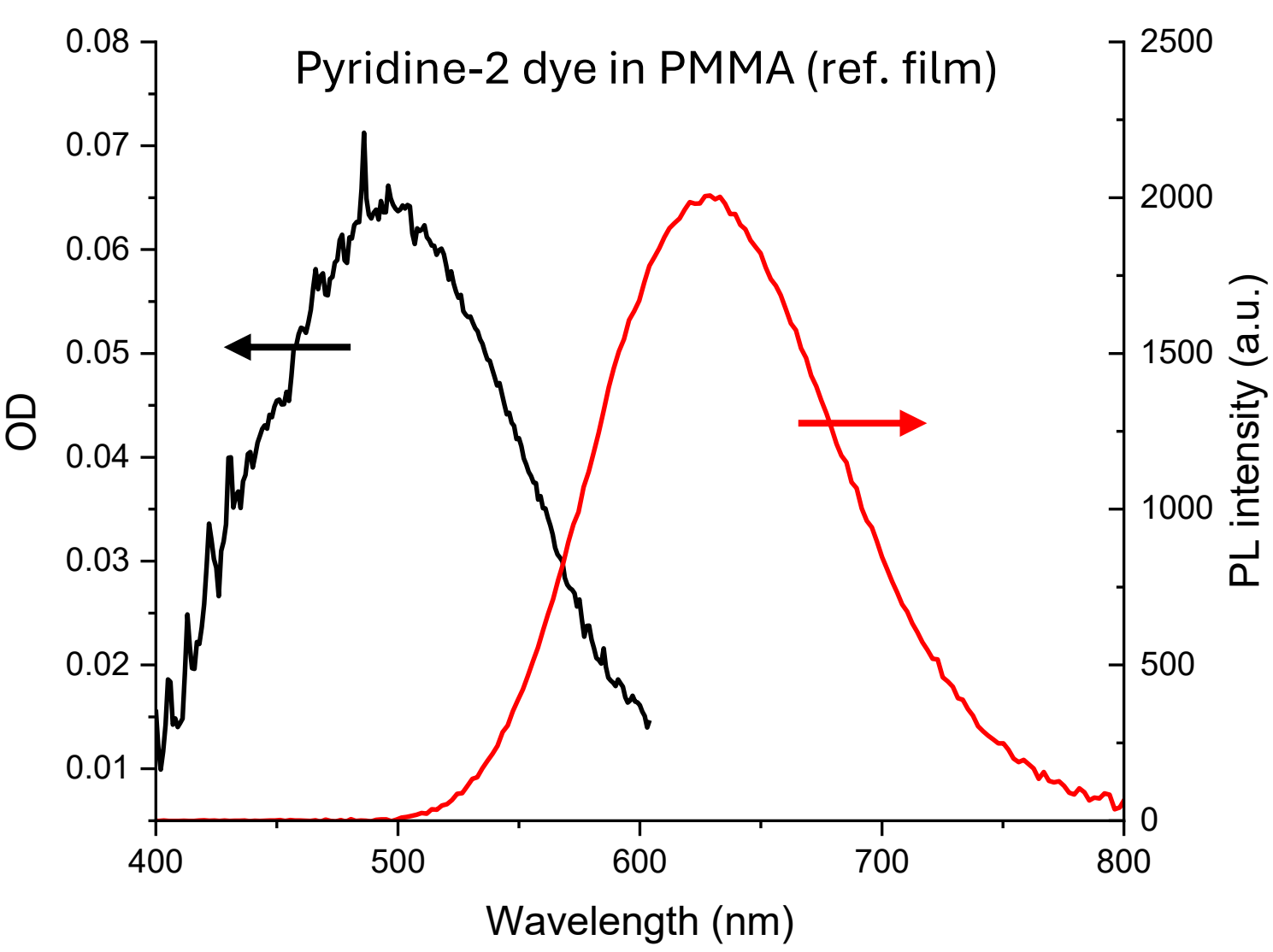

**Figure S1.** Absorption and PL spectra of reference dye film (pyridine-2 in PMMA matrix)

For the preparation of the dye-doped film we partially followed the procedures described in the literature.[1] Pyridine-2 dye was dissolved in DMSO at high concentration, the solution was kept at 100°C for 15 min to ensure complete dissolution. PMMA was separately dissolved in DMSO at a concentration close to saturation by heating the mixture at 100°C for 15 min. The two solutions were then combined in a 1:3 or 1:1 volumetric ratio, mixed thoroughly, and drop-cast onto plasma-treated hydrophilic glass substrates. The subsequent drying on a hot plate at 100 °C for 10 min was done to remove residual solvent before cooling to ambient conditions. As a result, smooth films of Pyridine-2 dispersed in PMMA with OD of the order of 0.1 were obtained for subsequent usage as reference samples (see absorption and PL spectra of the reference film in Figure S1).

## 2. Experimental setup

We used the same experimental setup as described in our previous publications.[2–5] In short, it is a home-built wide-field photoluminescence microscope system based on Olympus IX71 microscope. The sample can be excited by a 482 nm diode laser (PicoQuant, with a 200 ps pulse width) through a dry 40X objective lens (NA=0.6). The PL signal is then detected either by EM CCD (Princeton Instruments ProEM 512B) for measuring PLQY(*R, F*) map, or by a hybrid photomultiplier detector (PMA Hybrid- 42 PicoQuant) to measure TRPL decay kinetics. The hybrid detector is connected to a time-correlated single photon counting (TCSPC) module (PicoHarp 300 PicoQuant) for recording the arrival time of photons. The instrumental response function of our TCSPC setup is approximately 200 ps.

PLQY of reference samples was measured using the integrating sphere from Horiba scientific (Quanta ɸ) and the same laser for excitation. Measurements of the absorption spectra (including measurements in integrating sphere) were carried out using an UV-VIS-NIR spectrometer (Lambda 1050) from Perkin Elemer.

## 3. Measurement of PLQY(*R, F*) maps and TRPL decays

### 3.1 Estimation of the absorbance and the concentration of photogenerated charge carriers

For measuring PLQY(*R, F*) map, we used 68 combinations of the pulse fluence and repetition rate of the laser. For this we used 5 different pulse fluences, namely $F_1$ = 3.4x10$^{8}$, $F_2$ = 4x10$^{9}$, $F_3$ = 4.3x10$^{10}$, $F_4$ = 5.4x10$^{11}$, $F_5$ = 4.9x10$^{12}$ photons/cm$^2$ and several repetition rates ranging from 1 kHz to 80 MHz with a step of approximately 2-3 times (see Table S1 and S2). The value of the initial carrier concentration was estimated by assuming that each absorbed photon creates one electron – hole pair. Therefore, the initial carrier concentration ( $\Delta n_0 = \Delta p_0$) can be estimated in from the pulse fluence and absorptance of the sample using the following expressions:

$$\Delta n_0 = F_i \cdot \frac{A}{d} \tag{S3.1}$$

$$A = \frac{(I_0 - I_T - I_R)}{I_0} = (1 - T - R^*) \tag{S3.2}$$

Here, $F_i$ is the laser fluence, *A* is absorptance, and *d* is the sample thickness. $I_0$ is the intensity of the incident light, $I_T$ is the intensity of the transmitted light, and $I_R$ is the intensity of the reflected light. T is the transmission coefficient, $R^*$ is the reflection coefficient (reflection coefficient in this equation is represented by $R^*$ to differentiate it from the repetition rate $R$). The first equation assumes a uniform distribution of charge carriers over the sample thickness, and the second equation assumes negligible scattering (this is confirmed by measurements of the TC film absorption, scattering and reflections using integrating sphere).

The values of different pulse fluences and their corresponding initial carrier concentrations calculated using equation S3.1 for the sample studied here are listed in Table S1.

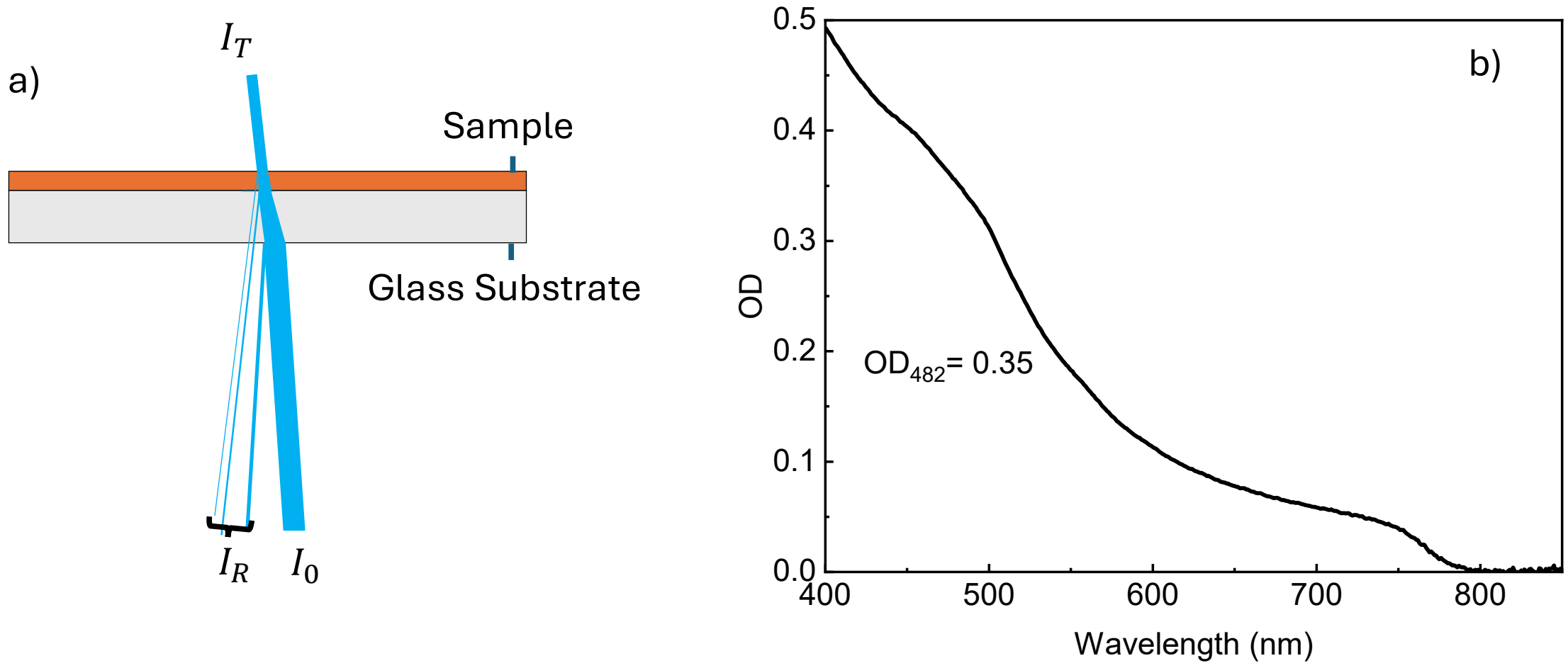

**Figure S2.** a) Schematic ray diagram for estimating the absorbance of the laser excitation light by the sample. The intensity of the incoming ($I_0$), reflected ($I_R$) and transmitted ($I_T$) light is measured with a power meter. b) absorption spectrum of the TC perovskite sample measured using integrating sphere.

Figure S2a illustrates the measurement of absorptance and reflectance using the excitation laser beam and a power meter. Results of multiple measurements on our 50 nm thin triple cation sample gave average transmittance 0.25 and average reflectance at the laser excitation wavelength 0.17. Which results in the average absorptance 0.58.

Figure S2b shows spectrum of the optical density of a triple cation sample of 50 nm film thickness measured in the integrating sphere. From the optical density at 482 nm ($OD$ = 0.35), we can find the value of the absorptance ($A$= $1\text{-}10^{-OD}$) to be 0.55, which matches very well with measured absorptance using the power meter (Figure S2a). It means that the scattering of the excitation light in the sample is quite small and can be ignored.

During the real PL measurements on the TC sample at the microscope, the sample was excited locally by a small beam (30 micrometers in diameter). The local absorptance could be slightly different due to inhomogeneity of the sample. However, in this experimental configuration we could measure incident $I_0$ and transmitted $I_T$ light but not reflected light $I_R$. However, the reflection coefficient is not very sensitive to the optical density of the sample, that is why it can be taken from the measurements with a broad beam like shown in Figure S2a. This allowed us to rather accurately estimate the amount of excitation light absorbed by different films for spectroscopy measurements as well as for the absolute PLQY calibration of the microscope (see SI section 4) using fixed reflection coefficient and still taking into account variable local optical density.

From the absorbance measurements we can estimate the absorption coefficient ($\alpha$) using the following relationship between absorptance and the absorption coefficient.

$$A = \frac{I_{absorbed}}{I_0} = (1 - S) \cdot (1 - R^*) \cdot (1 - e^{-\alpha d}) \quad \text{(S3.3)}$$

Using the values of the absorptance (0.58), scattering coefficient ($S$ = 0), the reflection coefficient ($R^*$ = 0.17), and film thickness (50 nm), we estimated the value of the absorption coefficient of the TC film at 482 nm to be $\alpha = 2.40 \times 10^5$ cm$^{-1}$.

**Supplementary Table S1**: Pulse fluences used in the study and the corresponding initial concentration of photogenerated charge carriers

| Label | Pulse fluence '*F*' (photons/cm$^2$) | Pulse fluence '*F*' (nJ/cm$^2$) | Photogenerated charge carrier density $\Delta n_0$ (cm$^{-3}$) |
|---|---|---|---|
| $F_1$ | $3.4 \times 10^8$ | 0.14 | $3.94 \times 10^{13}$ |
| $F_2$ | $4.0 \times 10^9$ | 1.64 | $4.64 \times 10^{14}$ |
| $F_3$ | $4.3 \times 10^{10}$ | 17.7 | $4.99 \times 10^{15}$ |
| $F_4$ | $5.4 \times 10^{11}$ | 220 | $6.26 \times 10^{16}$ |
| $F_5$ | $4.9 \times 10^{12}$ | 2000 | $5.68 \times 10^{17}$ |

### 3.2 Automated measurement of PLQY(*R, F*) maps

To measure the full PLQY(*R*, *F*) map, a home-developed LabVIEW program was used to control all parts of the setup.[6] This software was controlling the repetition rate of the laser, neutral density filters for controlling the excitation pulse fluence, neutral density filters in front of the charge coupled device (CCD) for controlling the amount of light that CCD receives. It also controlled the CCD camera (changing the integration time for taking a PL images and saving the captured images). All the required experimental conditions were uploaded to the software in the form of a table. Each row of the table sets the condition for one measurement.

For recording PL kinetics, the repetition rate of the laser was fixed at 10 kHz (close to the single pulse excitation regime for all $F_i$ while pulse fluences from $F_2$ to $F_5$ were used to excite the sample.

An example of the table, which was used to measure PLQY(*R, F*) map for the triple cation sample is shown in Table S2. Rows highlighted with light green color are data points for mapping PLQY, while the rows highlighted in light red color are the reference points for tracking the stability of the sample (always measured at the same condition). The evolution of the reference point is shown in Figure S4.

**Supplementary Table S2**: List of parameters used in the LabVIEW program to execute PLQY(*R, F*) mapping: M.N.- measurement number, Ex. OD – label of the excitation filter (≈optical density), Em. OD - label of the emission filter (≈optical density), Exp. time - exposure time of the CCD, Rep. rate - laser pulse repetition rate.

| M.N. | Ex. OD | Em. OD | Exp. Time (ms) | Rep. rate *R* (Hz) | M. N. | Ex. OD | Em. OD | Exp. Time (ms) | Rep. rate *R* (Hz) | M. N. | Ex. OD | Em. OD | Exp. Time (ms) | Rep. rate *R* (Hz) |
|---|---|---|---|---|---|---|---|---|---|---|---|---|---|---|
| 1 | 3 | 0 | 100 | 20000000 | 51 | 3 | 0 | 100 | 20000000 | 101 | 3 | 0 | 100 | 20000000 |
| 2 | 4 | 0 | 400000 | 200000 | 52 | 2 | 0 | 10000 | 20000 | 102 | 1 | 3 | 200 | 20000000 |
| 3 | 3 | 0 | 100 | 20000000 | 53 | 3 | 0 | 100 | 20000000 | 103 | 3 | 0 | 100 | 20000000 |
| 4 | 4 | 0 | 150000 | 500000 | 54 | 2 | 0 | 4000 | 50000 | 104 | 1 | 3 | 100 | 40000000 |
| 5 | 3 | 0 | 100 | 20000000 | 55 | 3 | 0 | 100 | 20000000 | 105 | 3 | 0 | 100 | 20000000 |
| 6 | 4 | 0 | 54000 | 1000000 | 56 | 2 | 0 | 1500 | 100000 | 106 | 1 | 3 | 100 | 80000000 |
| 7 | 3 | 0 | 100 | 20000000 | 57 | 3 | 0 | 100 | 20000000 | 107 | 3 | 0 | 100 | 20000000 |
| 8 | 4 | 0 | 18000 | 2000000 | 58 | 2 | 0 | 500 | 200000 | 108 | 0 | 0 | 1500 | 1000 |
| 9 | 3 | 0 | 100 | 20000000 | 59 | 3 | 0 | 100 | 20000000 | 109 | 3 | 0 | 100 | 20000000 |
| 10 | 4 | 0 | 6000 | 5000000 | 60 | 2 | 0 | 200 | 500000 | 110 | 0 | 0 | 500 | 2000 |
| 11 | 3 | 0 | 100 | 20000000 | 61 | 3 | 0 | 100 | 20000000 | 111 | 3 | 0 | 100 | 20000000 |
| 12 | 4 | 0 | 2000 | 10000000 | 62 | 2 | 0 | 100 | 1000000 | 112 | 0 | 0 | 200 | 5000 |
| 13 | 3 | 0 | 100 | 20000000 | 63 | 3 | 0 | 100 | 20000000 | 113 | 3 | 0 | 100 | 20000000 |
| 14 | 4 | 0 | 600 | 20000000 | 64 | 2 | 1 | 500 | 2000000 | 114 | 0 | 0 | 100 | 10000 |
| 15 | 3 | 0 | 100 | 20000000 | 65 | 3 | 0 | 100 | 20000000 | 115 | 3 | 0 | 100 | 20000000 |
| 16 | 4 | 0 | 200 | 40000000 | 66 | 2 | 1 | 200 | 5000000 | 116 | 0 | 1 | 500 | 20000 |
| 17 | 3 | 0 | 100 | 20000000 | 67 | 3 | 0 | 100 | 20000000 | 117 | 3 | 0 | 100 | 20000000 |
| 18 | 4 | 0 | 100 | 80000000 | 68 | 2 | 1 | 100 | 10000000 | 118 | 0 | 1 | 200 | 50000 |
| 19 | 3 | 0 | 100 | 20000000 | 69 | 3 | 0 | 100 | 20000000 | 119 | 3 | 0 | 100 | 20000000 |
| 20 | 3 | 0 | 300000 | 10000 | 70 | 2 | 2 | 500 | 20000000 | 120 | 0 | 1 | 100 | 100000 |
| 21 | 3 | 0 | 100 | 20000000 | 71 | 3 | 0 | 100 | 20000000 | 121 | 3 | 0 | 100 | 20000000 |
| 22 | 3 | 0 | 150000 | 20000 | 72 | 2 | 2 | 200 | 40000000 | 122 | 0 | 2 | 500 | 200000 |
| 23 | 3 | 0 | 100 | 20000000 | 73 | 3 | 0 | 100 | 20000000 | 123 | 3 | 0 | 100 | 20000000 |
| 24 | 3 | 0 | 60000 | 50000 | 74 | 2 | 2 | 100 | 80000000 | 124 | 0 | 2 | 200 | 500000 |
| 25 | 3 | 0 | 100 | 20000000 | 75 | 3 | 0 | 100 | 20000000 | 125 | 3 | 0 | 100 | 20000000 |
| 26 | 3 | 0 | 25000 | 100000 | 76 | 1 | 0 | 12000 | 1000 | 126 | 0 | 2 | 100 | 1000000 |
| 27 | 3 | 0 | 100 | 20000000 | 77 | 3 | 0 | 100 | 20000000 | 127 | 3 | 0 | 100 | 20000000 |
| 28 | 3 | 0 | 10000 | 200000 | 78 | 1 | 0 | 4500 | 2000 | 128 | 0 | 3 | 500 | 2000000 |
| 29 | 3 | 0 | 100 | 20000000 | 79 | 3 | 0 | 100 | 20000000 | 129 | 3 | 0 | 100 | 20000000 |
| 30 | 3 | 0 | 6000 | 500000 | 80 | 1 | 0 | 1500 | 5000 | 130 | 0 | 3 | 200 | 5000000 |
| 31 | 3 | 0 | 100 | 20000000 | 81 | 3 | 0 | 100 | 20000000 | 131 | 3 | 0 | 100 | 20000000 |
| 32 | 3 | 0 | 3000 | 1000000 | 82 | 1 | 0 | 500 | 10000 | 132 | 0 | 3 | 100 | 10000000 |
| 33 | 3 | 0 | 100 | 20000000 | 83 | 3 | 0 | 100 | 20000000 | 133 | 3 | 0 | 100 | 20000000 |
| 34 | 3 | 0 | 1000 | 2000000 | 84 | 1 | 0 | 200 | 20000 | 134 | 0 | 4 | 500 | 20000000 |
| 35 | 3 | 0 | 100 | 20000000 | 85 | 3 | 0 | 100 | 20000000 | 135 | 3 | 0 | 100 | 20000000 |
| 36 | 3 | 0 | 500 | 5000000 | 86 | 1 | 0 | 100 | 50000 | 136 | 0 | 4 | 200 | 40000000 |
| 37 | 3 | 0 | 100 | 20000000 | 87 | 3 | 0 | 100 | 20000000 | 137 | 3 | 0 | 100 | 20000000 |
| 38 | 3 | 0 | 200 | 10000000 | 88 | 1 | 1 | 500 | 100000 | 138 | 0 | 4 | 100 | 80000000 |
| 39 | 3 | 0 | 100 | 20000000 | 89 | 3 | 0 | 100 | 20000000 | 139 | 3 | 0 | 100 | 20000000 |
| 40 | 3 | 0 | 100 | 20000000 | 90 | 1 | 1 | 200 | 200000 | | | | | |
| 41 | 3 | 0 | 100 | 20000000 | 91 | 3 | 0 | 100 | 20000000 | | | | | |
| 42 | 3 | 0 | 100 | 40000000 | 92 | 1 | 1 | 100 | 500000 | | | | | |
| 43 | 3 | 0 | 100 | 20000000 | 93 | 3 | 0 | 100 | 20000000 | | | | | |
| 44 | 3 | 0 | 100 | 80000000 | 94 | 1 | 2 | 500 | 1000000 | | | | | |
| 45 | 3 | 0 | 100 | 20000000 | 95 | 3 | 0 | 100 | 20000000 | | | | | |
| 46 | 2 | 0 | 200000 | 2000 | 96 | 1 | 2 | 200 | 2000000 | | | | | |
| 47 | 3 | 0 | 100 | 20000000 | 97 | 3 | 0 | 100 | 20000000 | | | | | |
| 48 | 2 | 0 | 80000 | 5000 | 98 | 1 | 2 | 100 | 5000000 | | | | | |
| 49 | 3 | 0 | 100 | 20000000 | 99 | 3 | 0 | 100 | 20000000 | | | | | |
| 50 | 2 | 0 | 30000 | 10000 | 100 | 1 | 3 | 400 | 10000000 | | | | | |

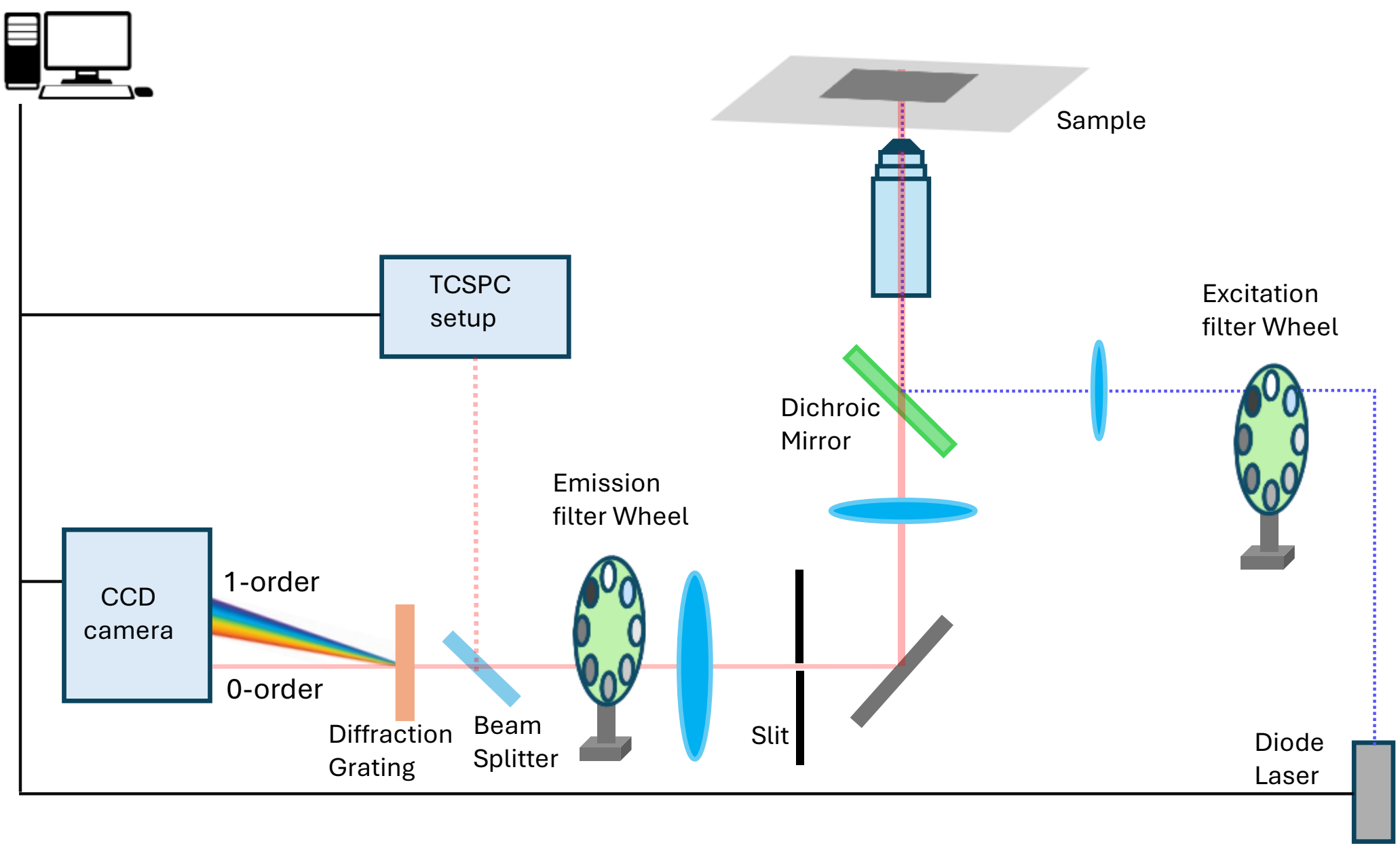

**Figure S3.** Schematic diagram of the wide-field fluorescence microscope used for the PLQY(*R, F*) mapping and TRPL decay measurements. The slit has an adjustable width, and it is fully opened for PLQY(*R, F*) measurements. The diffraction grating is placed in only for spectral measurements, the beam splitter is placed only for the TRPL measurements. PLQY(*R, F*) is acquired automatically according to the instruction list loaded to the controlling software.

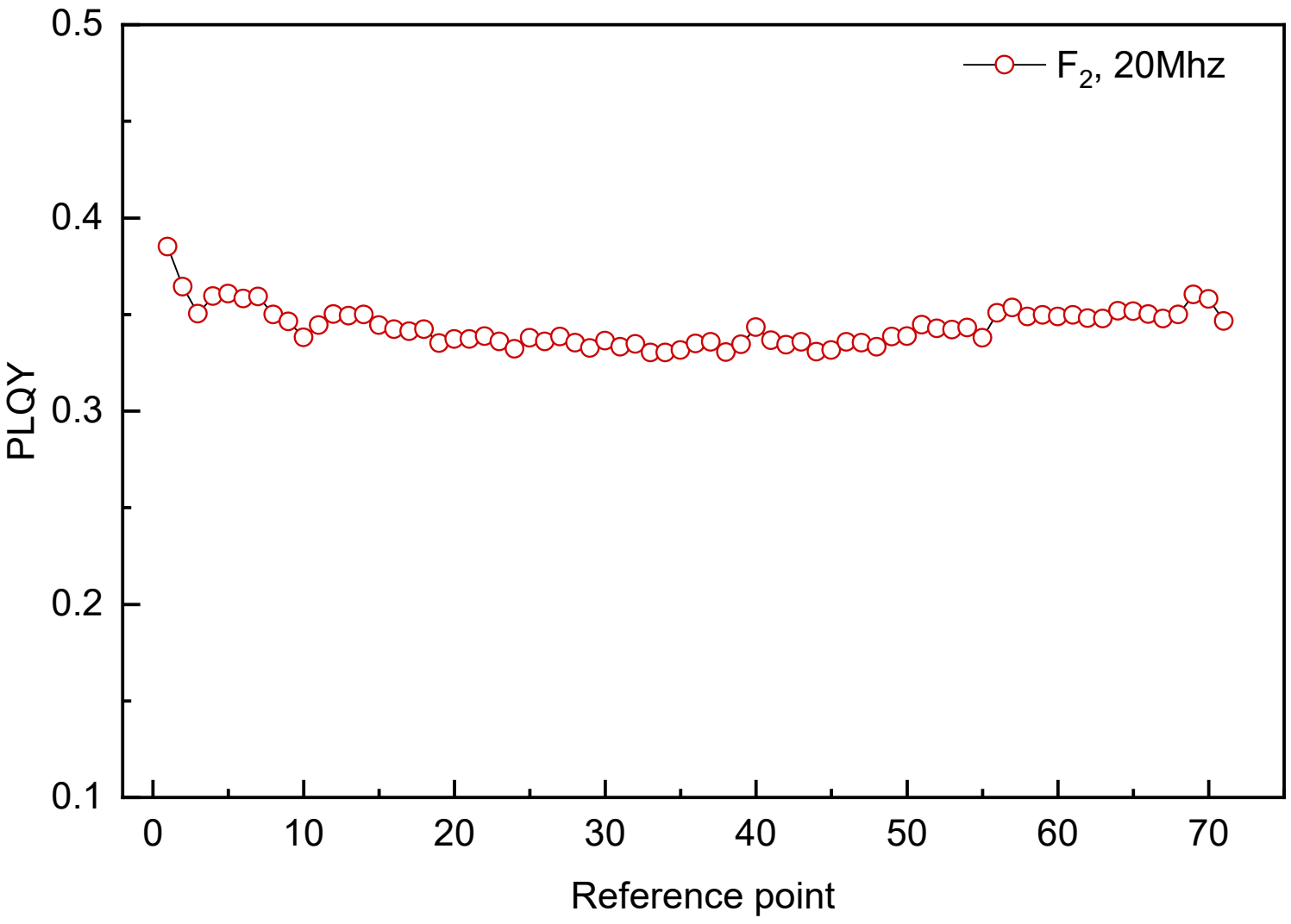

**Figure S4.** Time evolution of the PLQY of TC film at the reference point (PLQY @ $F_2$ 20MHz) over the whole PLQY(*R, F*) map measurement.

## 4. Calibration of the setup for absolute PLQY measurements

### 4.1. Measurements of PLQY of reference films using integrating sphere.

We used the integrating sphere and a fiber spectrometer (Avantes) to measure PLQY of the reference dye-doped PMMA films (see SI section 1.2) deposited to the same glass substrate that is used for the perovskite films. For excitation we used the same laser as for excitation of PL in the microscope. The size of the excitation spot was around 3 mm on the sample. The sample was placed against the bottom wall of the sphere. The absorption coefficient at 482 nm of the samples in this geometry in the sphere were in the range 0.15 – 0.2. The values of the measured PLQY were in the range of 0.3 – 0.4 depending on the sample.

### 4.2. Why can't we measure PLQY of the perovskite sample using the integrating sphere and use this value for calibration of the microscope?

In principle, we could just measure PLQY of the perovskite film using the integrating sphere in the same way as we did it for the reference dye film. The advantage would be that this reference sample is exactly the same in terms of its thickness and refractive index as the sample we investigated using the PL microscope setup. However, there are two problems causing rejection of this method:

1) PLQY of perovskites depends on the excitation power density. Therefore, to use this value to calibrate PLQY(*R, F*) maps we would need to know the excitation power density. We need to calibrate excitation density in the microscope and in the relative accuracy <10%, which is technically rather difficult. Moreover, the excitation spot has a distribution of the power density, which is practically not possible to take into account in the sphere. So, the measured PLQY is then averaged over a range of excitation conditions. Note that when measuring PLQY(*R, F*) map using the PL microscope we take the signal from the central part of the excitation spot only, where the excitation power is rather constant;

2) Perovskite samples are not sufficiently stable to use them for calibration. PL intensity can easily change by tens of % depending on light irradiation conditions;

3) Perovskite samples are not sufficiently homogeneous at the scale of the laser spot in the integrating sphere (in the microscope our excitation spot is 30 µm only).
Because we are aiming to have accuracy of the overall calibration <30%, the listed problems make this approach inappropriate.

### 4.3. The general principle of PLQY calibration

The main idea of the PLQY measurements of the perovskite films using the PL microscope is the following:

1) We have a reference sample (the thin polymer film doped with a fluorescent dye on a glass substrate). This sample we can image via PL microscope and also measure its PLQY using integrating sphere. The sample should be stable. PLQY of the reference sample should not depend on the excitation power density, which is generally true to diluted dye solutions as well as for dye molecules well dispersed in polymer matrices.

2) If the geometry of the reference sample (film thickness, substrate, refractive index *etc.*) and the actual sample of interest (the perovskite films in our case) are the same (or at least very close), then the ratio of PL intensities detected by the CCD from the sample and reference sample should be equal to the ratio of their PLQYs if all other conditions (laser excitation power, acquisition time, local absorption coefficient, setup sensitivity) are the same.
If the mentioned conditions are not the same, the differences should be taken into account. The most difficult issue to take into account is the wavelength-dependent sensitivity of the setup. Below it will be described how it was measured in our case.

PL intensity of the reference sample excited by laser power *P* (mW) is acquired over the exposure time $\tau$ (s) under specific settings of the CCD (gain, readout rate etc.), specific objective lens and all other optical components between the sample and the CCD detector.
Number of counts $C_{ref}$ obtained by integration of the CCD camera image (after subtraction of the background) can be expressed as:

$$C_{ref} = \beta_{ref}\, P_{ref}\, \tau_{ref}\, \Phi_{ref} A_{ref} \tag{S4.1}$$

Where $\beta_{ref}$ – is the detection efficiency of the microscope for the PL of the reference sample, $A_{ref}$ is the local absorptance of the reference sample. The index **ref** hereafter designates specific laser power, exposure time and other parameters used for the reference sample.

The same equation is valid for the sample of interest, (index ***s*** hereafter is for the sample of interest):

$$C_s = \beta_s\, P_s\, \tau_s\, \Phi_s\, \mathrm{A}_s \tag{S4.2}$$

By division these two expressions on each other, we obtain:

$$\frac{C_s}{C_{ref}} = \frac{\beta_s\, P_s\, \tau_s\, \Phi_s A_s}{\beta_{ref}\, P_{ref}\, \tau_{ref}\, \Phi_{ref} A_{ref}} = \left(\frac{\beta_s}{\beta_{ref}}\right) \cdot \frac{P_s\, \tau_s\, \Phi_s A_s}{P_{ref}\, \tau_{ref}\, \Phi_{ref} A_{ref}} \tag{S4.3}$$

And for the sample PLQY we get:

$$\Phi_s =\cdot\, \Phi_{ref} \cdot \frac{P_{ref}\, \tau_{ref} A_{ref}}{C_{ref}} \cdot \left(\frac{\beta_{ref}}{\beta_s}\right) \frac{C_s}{P_s\, \tau_s A_s} = B\ \cdot \frac{C_s}{P_s\, \tau_s A_s} \tag{S4.4}$$

where,

$$B = \Phi_{ref} \frac{P_{ref}\, \tau_{ref} A_{ref}}{C_{ref}} \cdot \left(\frac{\beta_{ref}}{\beta_s}\right) \tag{S4.5}$$

Then, if the sensitivities of the setup for the reference sample and the sample of interest are the same $\beta_s = \beta_{ref}$ (this is the case when: i) the reference and the sample have identical emission spectrum, or ii) when the sensitivity of the setup does not depend on wavelength) we can use this equation directly to obtain the coefficient *B* because all parameters there can be easily measured experimentally. In reality, $\beta_s \neq \beta_{ref}$, thus we need to measure somehow the spectral sensitivity of the imaging mode of the microscope.

### 4.4. Measuring the spectral sensitivity of the setup

Here our task is to measure $\frac{\beta_s}{\beta_{ref}}$ and apply this coefficient to Eq. S4.5.Our microscope can be converted to a spectrometer by introducing a slit and a diffraction grating, as shown in Figure S3. The spectrum is then detected by the same CCD camera as for imaging.
By measuring the spectrum $M_{Lamp}(\lambda)$ of the calibration lamp (Avantes, AvaLight-HAL-CAL-Mini), for which actual spectrum is known from the producer ($K_{Lamp}(\lambda)$, in energy/nm) we can obtain the sensitivity coefficient of the microscope in the spectral mode:

$$c_{sensitivity}(\lambda) = \frac{\lambda \cdot K_{Lamp}(\lambda)}{M_{Lamp}(\lambda)} \tag{S4.6}$$

Here multiplication by λ is needed to convert the units of energy/nm of the tabulated lamp spectrum to photons/nm. Finally, the units of the $c_{sensitivity}$ *(λ)* function are: [(Photons/nm)/(CCD counts)]

Whether the CCD counts are proportional to the incident energy or the number of photons does not really matter here, because the actual units will be canceled when multiplied by the real spectrum *M(λ)* of sample of interest to obtain its corrected spectrum in [Photons/nm], see Figure (S5)'

$$S(\lambda) = M(\lambda) \cdot c_{sensitivity}(\lambda) \tag{S4.7}$$

It is important to realize, however, that this sensitivity curve is **for the spectral mode only** and cannot be yet directly applied to the imaging mode. This is because in the spectral mode the diffraction grating is present in the optical path adding its own wavelength-dependent spectral efficiency (the grating is absent in the imaging mode).
Now let's calibrate the imaging mode having the spectral model calibrated.

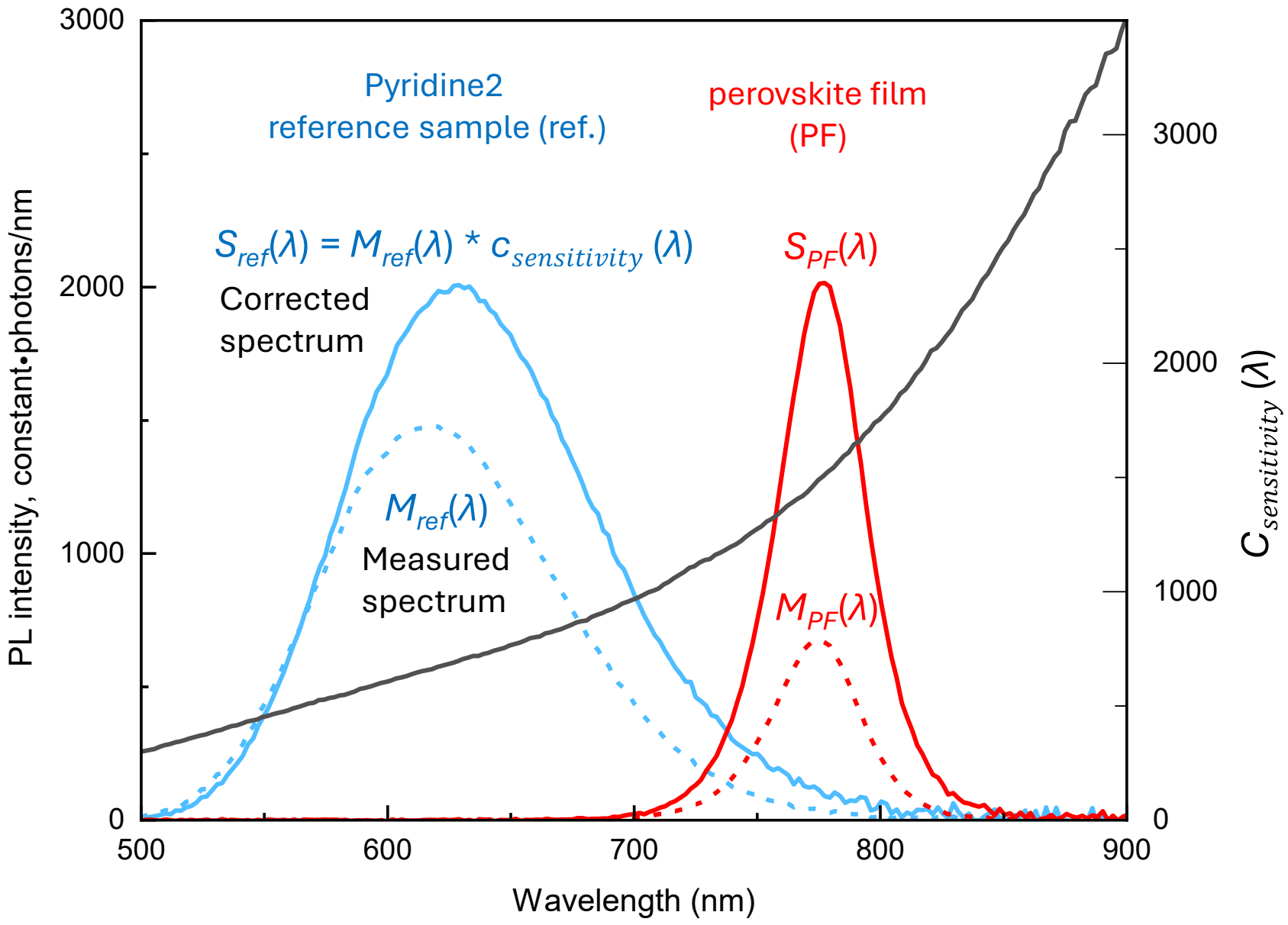

**Figure S5.** Dashed lines - spectra of the reference sample (pyridine2) and the TC perovskite film (PF) measured using spectral mode of the microscope (M). Solid lines - corrected spectra (S) using the sensitivity coefficient $c_{sensitivity}(\lambda)$ (black solid line).

We use two samples for this calibration:
i) **reference film of known PLQY** (called **ref.** hereafter), which was measured in the integrating sphere.
ii) **reference perovskite film** (called **PF** here after), which we use here to have a source of PL **with the same spectrum** as the actual perovskite sample PLQY(*R, F*) map is measured.
We measure under the same excitation power *P* and the same slit (in terms of its width and the position over the laser excitation spot) the following CCD images (see Figure S6):

1) an image of the **ref.** film,
2) an image of the **ref.** film with the slit,

(these two images will allow to measure the laser excitation power within the slit)

3) an image of the **ref.** film with the slit and with the grating,

(this measurement is needed to obtain the calibrated spectrum of the reference sample)

4) image of the **PF** sample with the slit,
5) image of the **PF** sample with the slit and with the grating,

(this measurement is needed to obtain the calibrated spectrum of the **PF** film).Notes about the **ref** sample: the sample should not be moved between measurement #1, #2 and #3, PL of the ref. sample should be stable over these measurements, PLQY should not depend on the excitation power density, the local absorption coefficient *A* should be measurable.

<u>Note about the **PF** sample:</u> the PL intensity of this sample needs to be stable only during measurements #4 and #5, which can be ensured even for perovskites;

Laser fluence is '$F$' for all measurements

PL image of the reference film, no slit, no grating:

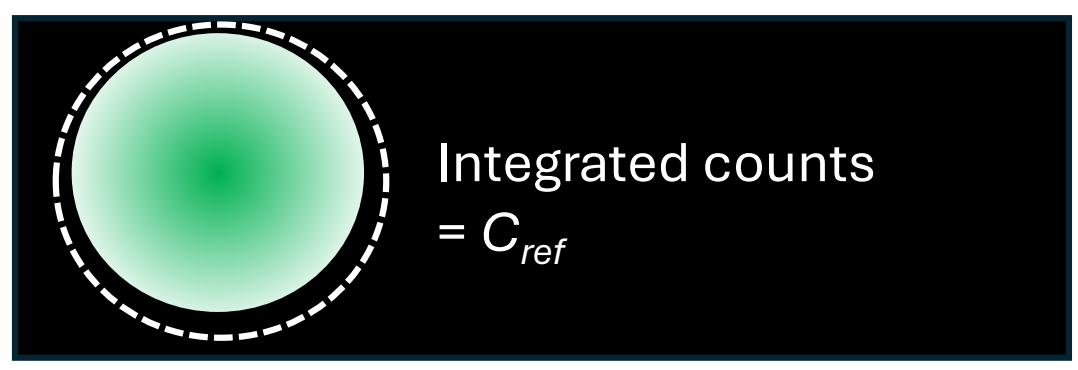

$$F_{slit} = F \cdot \frac{C_{ref_slit}}{C_{ref}}$$

Reference film with slit:

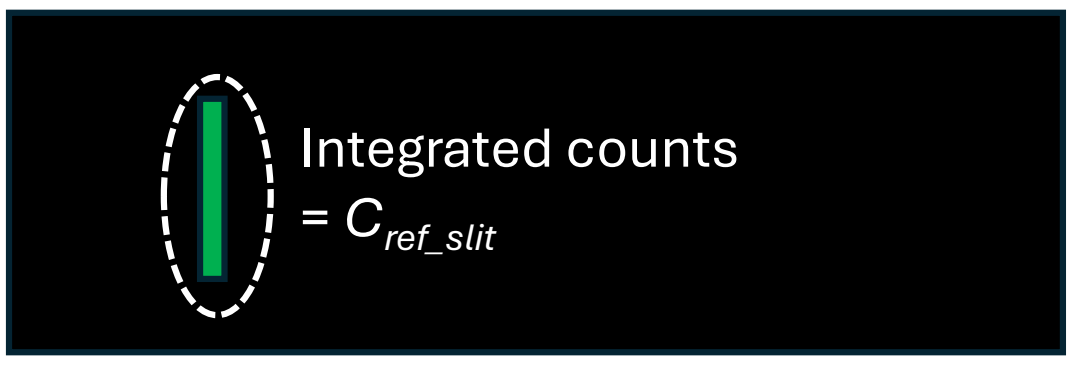

The slit width and its position relative to the excited area is the same for all further measurements

Reference film with slit and grating:

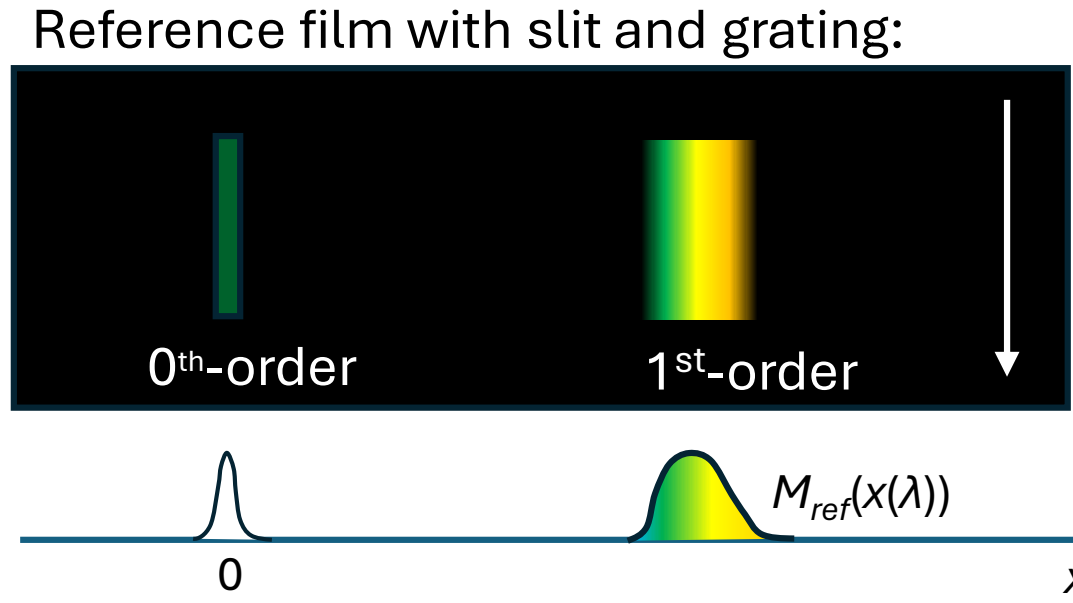

Integration over the pixel columns and converting pixels to wavelength gives measured spectrum $M_{ref}(\lambda)$, which is then converted to the calibrated spectrum
$S_{ref}(\lambda) = M_{ref}(\lambda)\ c_{sensitivity}(\lambda)$

PL image of the Reference Perovskite Film with the slit:

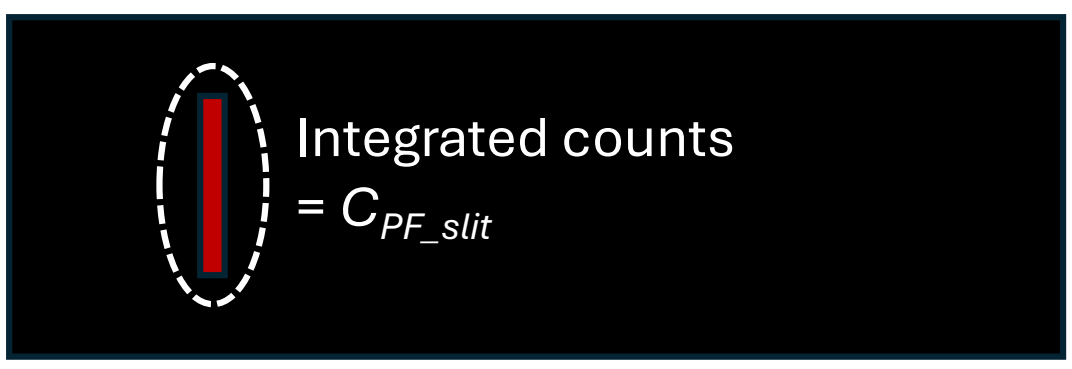

The Reference Perovskite Film with slit and grating:

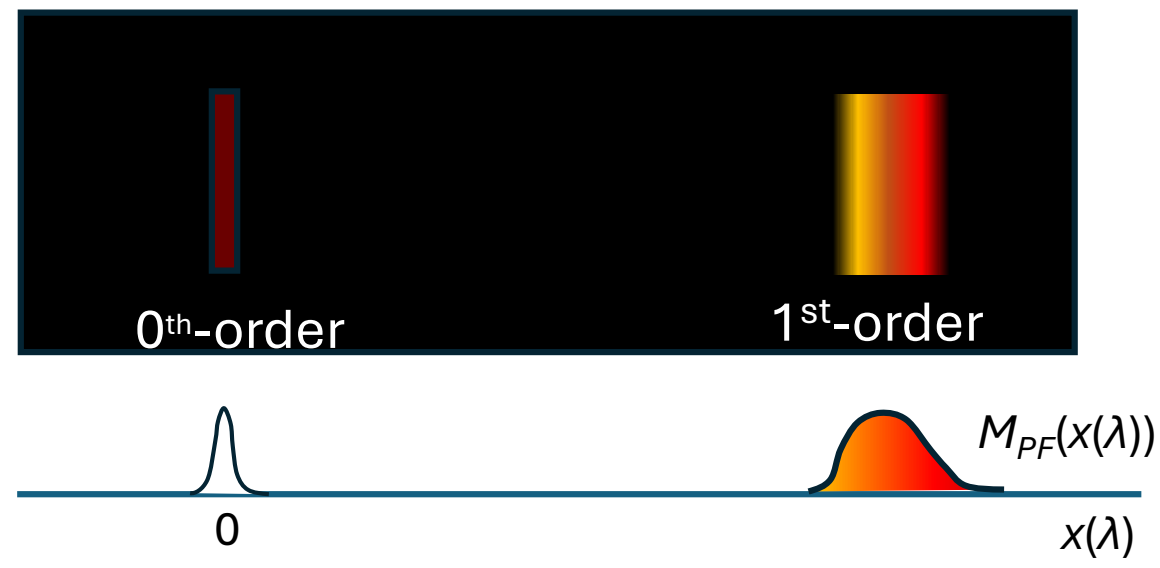

Integration over the pixel columns and converting pixels to wavelength gives measured spectrum $M_{PF}(\lambda)$, which is then converted to calibrated spectrum $S_{PF}(\lambda)$

**Figure S6.** Illustration of the measurements required for spectral sensitivity calibration of the imaging mode of the PL microscope. The pictures illustrate CCD images.

The parameters used below in the equations are also explained in Figures S5 and S6. The laser power hitting the excited area of the sample observed through the slit can be calculated as:

$$F_{slit} = F \cdot \frac{C_{ref_slit}}{C_{ref}} \tag{S4.8}$$

Where *P* is the total power of the laser hitting the sample, $C_{ref_slit}$ and $C_{ref}$ are the total counts on of the CCD sensor (sum of all counts minus background).
The PL intensity of the reference sample (ref) detected through the slit is:

$$C_{ref_slit} = \Phi_{ref} A_{ref} P_{slit} \beta_{ref} \tag{S4.9}$$

The PL intensity of the perovskite reference film (PF) detected through the slit is:

$$C_{PF_slit} = \Phi_{PF} A_{PF} P_{slit} \beta_{PF} \tag{S4.10}$$

Here the excitation conditions are the same in both cases.
The ratio of these signals is:

$$\frac{C_{PF_slit}}{C_{ref_slit}} = \frac{\Phi_{PF} A_{PF}}{\Phi_{ref} A_{ref}} \cdot \left(\frac{\beta_{PF}}{\beta_{ref}}\right) \tag{S4.11}$$

Now let us use the calibrated spectra of the ref. and PF samples obtained at the same excitation and detection conditions. Their ratio is equal to the ratio of the number of photons emitted by the ref. and PF samples:

$$\frac{\int S_{PF}(\lambda) d\lambda}{\int S_{ref}(\lambda) d\lambda} = \frac{\Phi_{PF} A_{PF} P_{slit}}{\Phi_{ref} A_{ref} P_{slit}} = \frac{\Phi_{PF} A_{PF}}{\Phi_{ref} A_{ref}} \tag{S4.12}$$

Combining the last two equations (S4.11 and S4.12) we obtain:

$$\frac{C_{PF_slit}}{C_{ref_slit}} = \frac{\int S_{PF}(\lambda) d\lambda}{\int S_{ref}(\lambda) d\lambda} \cdot \left(\frac{\beta_{PF}}{\beta_{ref}}\right) \tag{S4.13}$$

Or

$$\frac{\beta_{ref}}{\beta_{PF}} = \frac{\int S_{PF}(\lambda) d\lambda}{\int S_{ref}(\lambda) d\lambda} \cdot \frac{C_{ref_slit}}{C_{PF_slit}} \tag{S4.14}$$

Thus, we obtained the needed ratio of the sensitivities. Now we substitute it to Eq. S4.5 and assume that the sample has the same spectrum as the PF sample:

$$B = \Phi_{ref} \frac{P_{ref}\, \tau_{ref} A_{ref}}{C_{ref}} \cdot \frac{\int S_{PF}(\lambda) d\lambda}{\int S_{ref}(\lambda) d\lambda} \cdot \frac{C_{ref_slit}}{C_{PF_slit}} \tag{S4.15}$$

Because PL of the ref. sample is proportional to the excitation power we can write

$$C_{ref} = C_{ref_slit} \frac{P_{ref}}{P_{slit}}$$

This simplifies the equation to:

$$B = \Phi_{ref} \frac{F_{ref}\, \tau_{ref} A_{ref}}{C_{ref_{slit}} \frac{F_{ref}}{F_{slit}}} \cdot \frac{\int S_{PF}(\lambda) d\lambda}{\int S_{ref}(\lambda) d\lambda} \cdot \frac{C_{ref_{slit}}}{C_{PF_{slit}}} = \Phi_{ref} \frac{\tau_{ref} A_{ref} F_{slit}}{C_{PF_slit}} \cdot \frac{\int S_{PF}(\lambda) d\lambda}{\int S_{ref}(\lambda) d\lambda} \cdot$$

So,

$$B = \Phi_{ref} \frac{\tau_{ref} A_{ref} P_{slit}}{C_{PF_slit}} \cdot \frac{\int S_{PF}(\lambda) d\lambda}{\int S_{ref}(\lambda) d\lambda}. \quad (S4.16)$$

The units of the conversion factor ***B*** are $\left[\frac{mW \cdot ms}{counts}\right]$.

Note that for determination of the conversion factor ***B*** we do not need to know the absorption coefficient of the reference perovskite film (**PF**).

For realization of this calibration procedure, we used two reference samples (ref.) of Pyridine-2 dispersed in PMMA matrix deposited on glass. The samples are labelled S1 and S2 with PLQY equal to 0.45 and 0.38 and averaged local absorption coefficient 0.067 and 0.075 (measured at the microscope). TC perovskite sample (the same as used for the actual measurements) was used as the PF sample. It had absorption coefficient 0.58. The excitation light intensity for calibration had fluence $F_1$ and repetition rate 80 MHz. At this conditions PLQY of the TC sample was 0.22.

The statistics of *B* measured several times with two different reference samples S1 and S2 is shown in Figure S7 below.

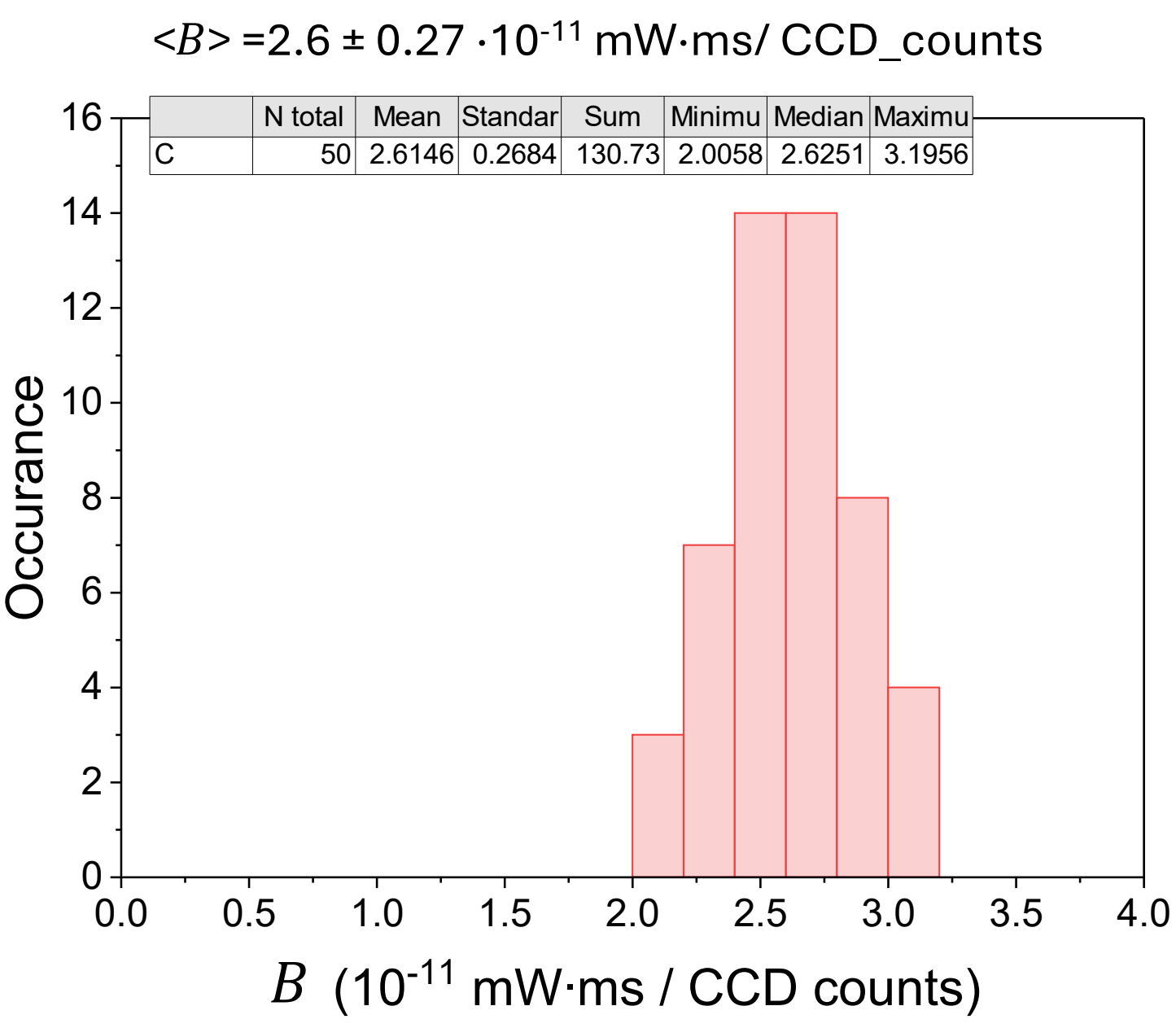

**Figure S7.** Statistics of the *B* values obtained from 50 measurements with two different reference dye samples of Pyridine-2 (S1, S2), the same TC perovskite sample was used for the spectral calibration.

## 4.5. Influence of different factors on the accuracy of the PLQY calibration (*B* coefficient)

### 4.5.1 Influence of inhomogeneous local absorption of the reference film

For accurate calibration we need to make sure that the PLQY of the dye reference sample does not depend on its local absorption. This is important because the PLQY measured inside the integrating sphere corresponds to the average absorbance over a large excitation spot (several mm). Whereas the PL signal measured on the microscope comes from a much smaller excitation spot (30 micrometers) where possible inhomogeneity of the sample cannot be ignored in general. The inhomogeneity of the sample however does not matter for use if the PLQY of the sample is independent of the number of absorbed photons (constant PLQY) – in this case variation of the PL signal is due to variation of the absorption coefficient only, which we always measure locally. To verify this, we measured the PL counts as a function of local absorbance and found that the PL counts are linearly proportional to it (Figure S8). This indicates that any change in local absorbance does not change the local PLQY and, therefore, any random local region is a representative of the PLQY of the whole sample. Note that this is what one would expect from dye molecules well-dispersed in a polymer matrix.

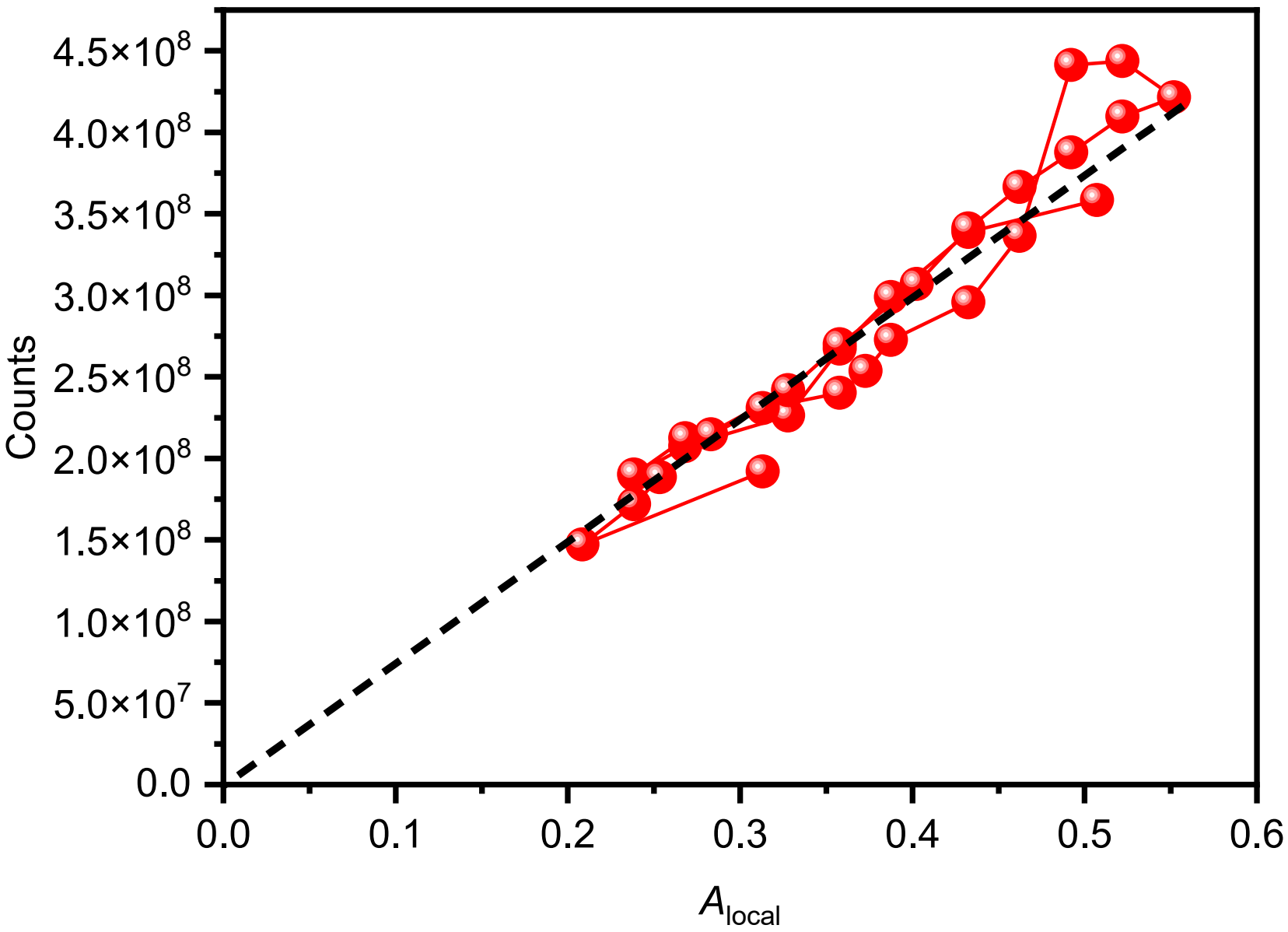

**Figure S8.** CCD counts (PL intensity) as a function of the local absorbance of the Pyridine-2 sample. The counts are linear to absorptance; however, some discrepancy is visible showing the level errors.

### 4.5.2 Influence of misfocusing of the microscope on the PL counts

The counts on the CCD are dependent on the focus of the microscope on the sample. In principle the focus position of the microscope used for its calibration and for the actual measurement should be exactly same. In practice it is not so easy to match the focus condition with accuracy better than 5 micrometers because samples are rather uniform and do not usually have clear features to focus on. To estimate the effect of mis-focusing on the integrated intensity detected by the CCD we intentionally move the microscope objective up and down in steps of 2 µm.
We observed that the total counts on the CCD do not change more than 1 – 3 % if we are within 5 -10 µm range from the ideal focus condition (Figure S9). We also noticed that deviating 5 -10 µm from the best focus condition changes the contrast of the image substantially (easily visible), therefore it is always possible to match the focus for the calibration and the actual measurement within this range.

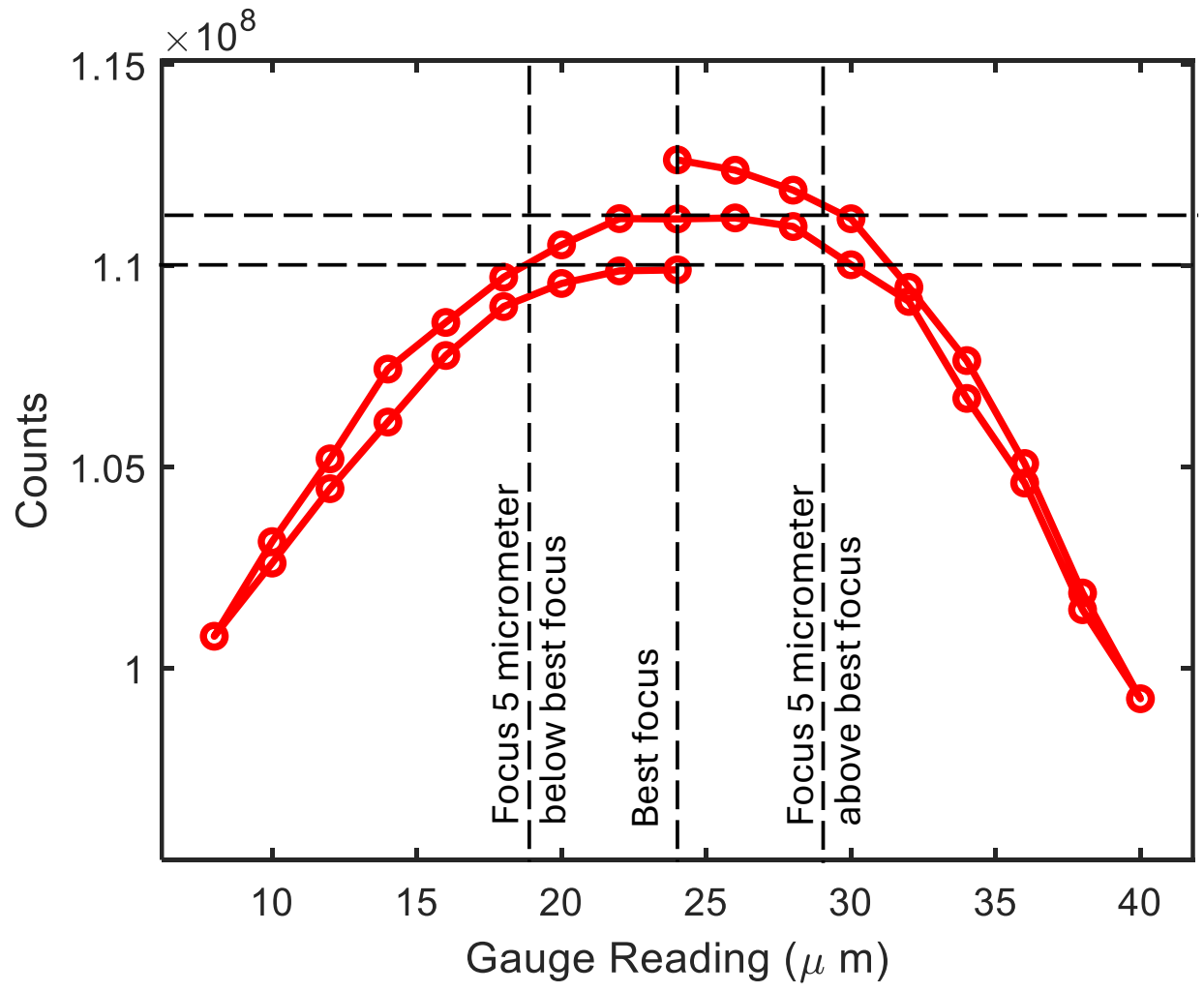

**Figure S9.** Dependence of the integrated counts at CCD (the whole excited spot is integrated; a reference dye film was used as a sample) on the shift of the focus (40X objective lens, NA = 0.6). Vertical dashed lines indicate a ±5 μm deviation in the reading of the microscope's focusing knob relative to the optimal focus condition. The hysteresis visible in the data shows the magnitude of PL bleaching of the reference film (see section 4.5.3).

#### 4.5.3 Influence of the instability of the reference film

As it has been mentioned, the reference dye film should be as stable as possible for accurate calibration. To our experience, any sample shows some bleaching if one excites strong enough and for long enough time. In our case the total exposure of the ref. sample to light was no more than a few seconds to accomplish all measurements described in Figure S6. The effect of a much longer series of measurements can be seen in Figure S9 where the same place of the reference sample was measured 33 times. The hysteresis visible at the graph shows bleaching, it was around 2% only. So, the reference films used in this study were stable enough for calibration.

#### 4.5.4 Influence of the instability of the perovskite reference film

We do not use the perovskite film as a real reference for absolute PL intensity due to an unpredictable behavior of its PLQY under light irradiation and other issues (see section 4.4). We use it for spectral calibration only where it has to be stable between just two measurements, each of them can last < 1s. One can even pre-bleach the sample in order to make its PL more stable. We do not see any problem using perovskites sample for the purposes described in section 4.4 despite their light sensitivity.

## 5. The initial TRPL intensity drop

## 5.1. Effect of charge carrier diffusion

According to the Beer-Lambert law (Figure S10a) the initial distribution of the photogenerated charge carrier concentration created by a laser pulse arriving at time $t$ = 0 is given by:

$$n(x, t = 0) = p(x, t = 0) = \gamma \cdot e^{-\alpha x} \qquad \text{(S5.1)}$$

Where x is the distance from the front surface of the film and $\gamma$ is a constant dependent on absorption coefficient $\alpha$ and excitation fluence.
Then due to charge carrier diffusion over the film thickness $d$ the distribution becomes homogenous with an average concentration $\bar{n}$ (Figure S10a). This average concentration can be connected to $\gamma$ by assuming that the total number of charge carriers do not change during the homogenization (the homogenization is much faster than recombination), using Eq. S5.1 we get:

$$\int_0^d n(x)dx = \gamma \int_0^d e^{-\alpha x} dx = \bar{n}d$$

$$\frac{\gamma}{\alpha}(1 - e^{-\alpha d}) = \bar{n}d$$

$$\bar{n} = \frac{\gamma}{\alpha d}(1 - e^{-\alpha d}) \qquad \text{(S5.2)}$$

Let's compare the initial PL intensity and the PL intensity after homogenization. Assuming the semiconductor is intrinsic, the initial PL intensity (per $cm^2$) is given by:

$$I_{PL}(0) = k_r \cdot \int_0^d n(x, t = 0)p(x, t = 0)dx = k_r \cdot \gamma^2 \int_0^d \exp(-2\alpha x)\, dx =$$

$$= \frac{k_r \cdot \gamma^2}{2\alpha}(1 - e^{-2\alpha d}) \quad \text{(S5.1)}$$

After homogenization at time $t'$ the intensity is:

$$I_{PL}(t') = k_r \int_0^d \bar{n}^2 dx = k_r \bar{n}^2 d = k_r\, d \left[\frac{\gamma}{\alpha d}(1 - e^{-\alpha d})\right]^2 = k_r \frac{\gamma^2}{d\alpha^2}(1 - e^{-\alpha d})^2 \quad \text{(S5.2)}$$

Thus, their ratio:

$$\frac{I_{PL}(t')}{I_{PL}(0)} = \frac{k_r \frac{\gamma^2}{d\alpha^2}(1 - e^{-\alpha d})^2}{\frac{k_r \cdot \gamma^2}{2\alpha}(1 - e^{-2\alpha d})} = \frac{2}{\alpha d}\frac{(1 - e^{-\alpha d})^2}{(1 - e^{-2\alpha d})} \quad \text{(S5.3)}$$

Then the initial relative PL intensity drop $\xi$ defined in Figure S10b and used in the main text for describing PL decays is given by:

$$\xi = 1 - \frac{I_{PL}(t')}{I_{PL}(0)} = 1 - \frac{2}{\alpha d}\frac{(1 - e^{-\alpha d})^2}{(1 - e^{-2\alpha d})} \quad \text{(S5.4)}$$

The initial PL drop for 50 nm thick film is plotted in Figure S10c as a function of the absorption coefficient of the material. For our case ($\alpha$=2.4x10$^5$ cm$^{-1}$) we get ξ≈ 0.1 only, while in the experiment get ξ≈ 0.9 (Figure S10d). Thus, it is not possible that the experimentally observed PL drop was caused by diffusion.

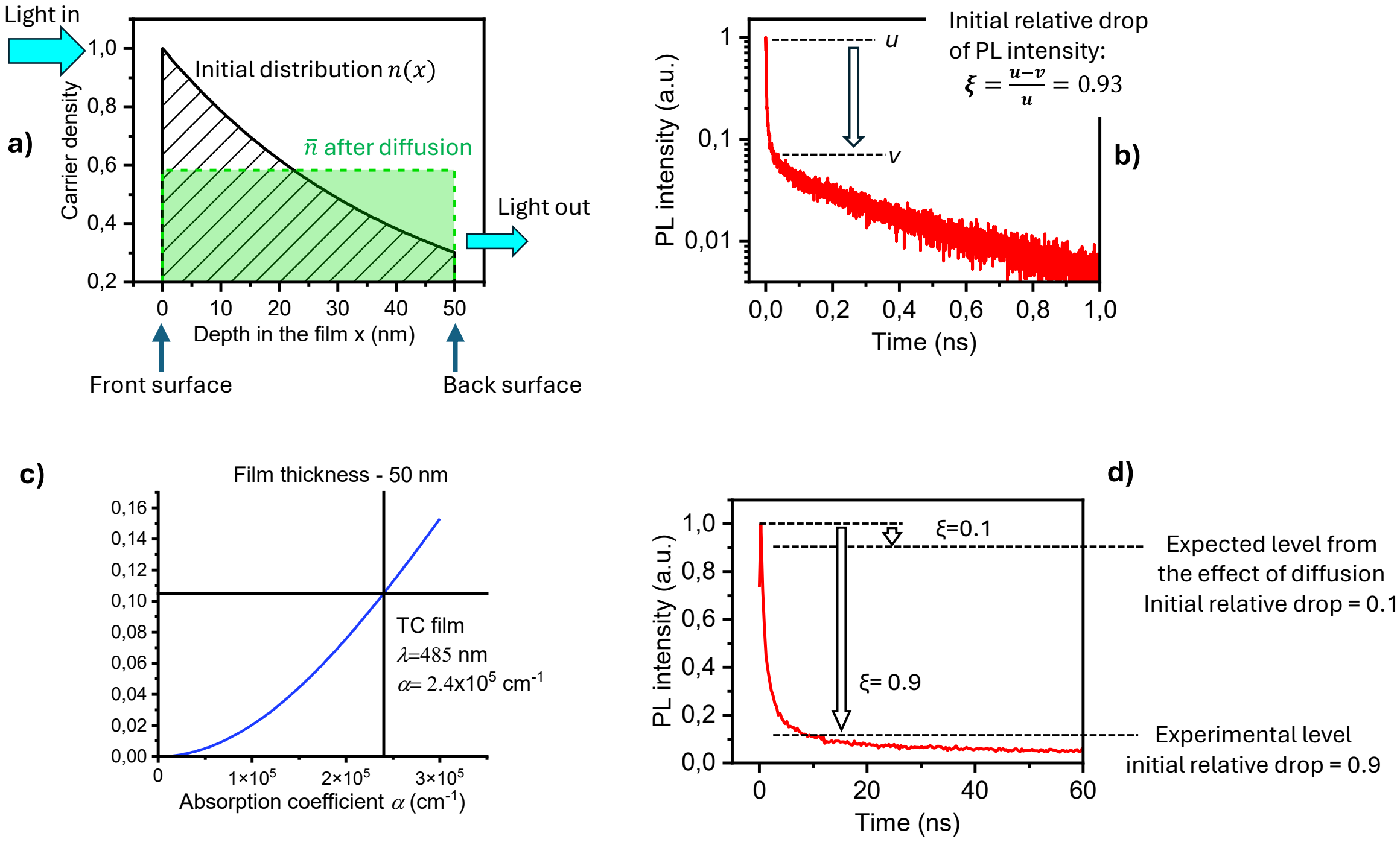

**Figure S10**. (a) Initial distribution of photogenerated charge carrier concentration over the film thickness according to Beer-Lambert law (black line) and the homogeneous distribution after charge carrier diffusion (green line) calculated for 50 nm thin TC film. (b) Experimentally measured PL decay of the TC film excited by $F_2$ pulse fluence, the initial relative intensity drop ξ is defined in the figure. (c) Relative PL intensity drop due to homogenization calculated by Eq. S5.4 for 50 nm thick film as a function of the absorption coefficient of the material. Absorption coefficient for TC is around 2.4x10$^5$ cm$^{-1}$ resulting in ξ=0.1. (d) comparison of the experimental relative PL intensity drop and the expected one from the effect of diffusion only.

### 5.2. Effect of trap saturation

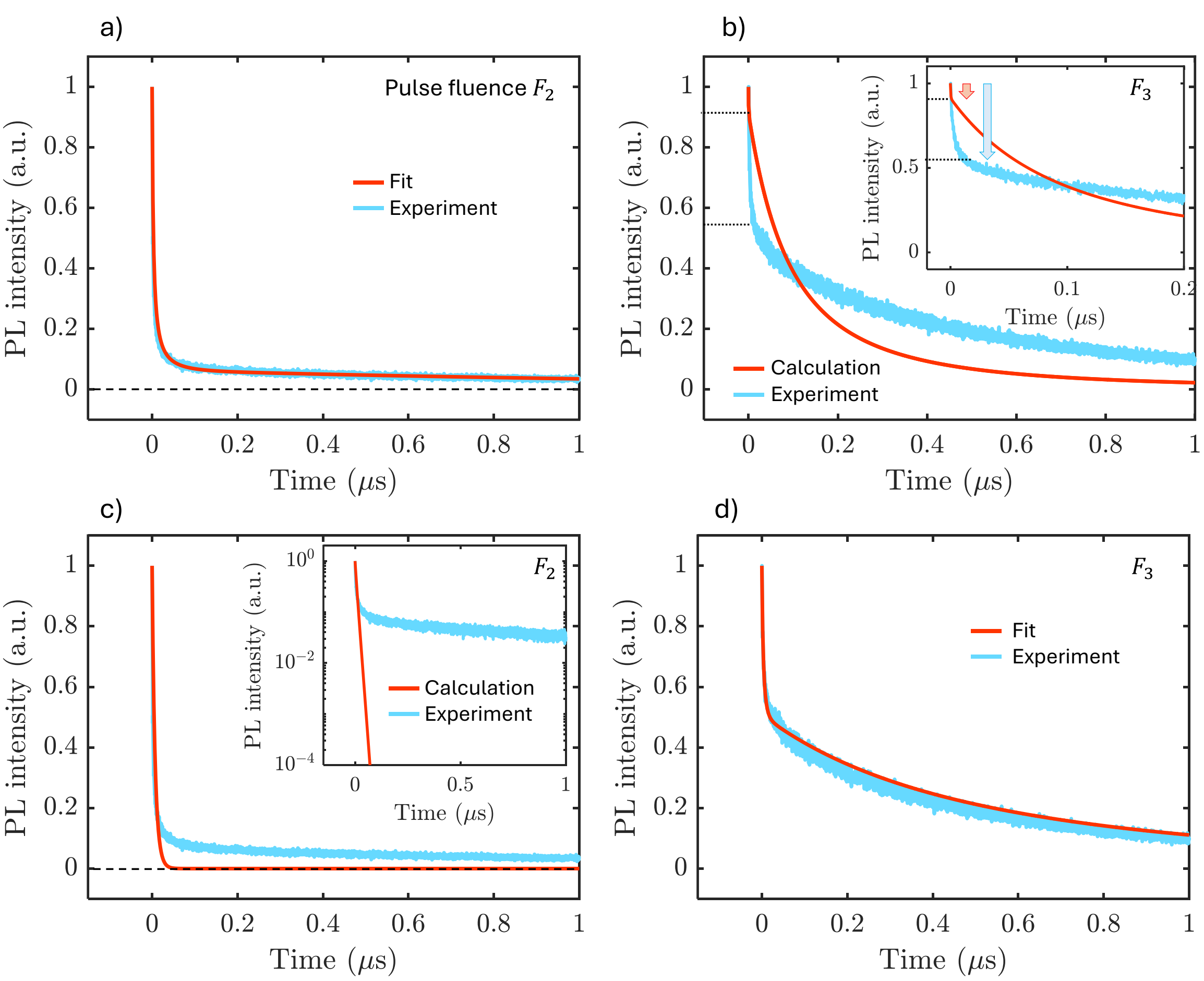

**Figure S11.** Attempts to explain the TRPL including the initial fast component using a saturable deep trap for fluence F2 (a) and calculate for fluence F3 (b) and to fit TRPL for F3 (d) and calculate it for F2 (c).

Figure S11a shows the TRPL decay at low fluence (*F2*) fitted using a model with a saturable deep trap (parameters $k_{t1}$, $N_{t1}$, $k_{n1}$ and $k_r$) hereafter referred to as SRH-0. However, this model with the same parameters does not reproduce the observed TRPL decay at an order of magnitude higher fluence (*F3*), leading to a very large mismatch (factor of 5) in the initial PL intensity drop and in the overall decay shape (Figure S11b). When calculation predicts the initial PL intensity drop to the level of 0.9, PL intensity drops to the level of 0.5 in experiment (5 times difference in the relative amplitude).

Using SRH-0 model with different parameters we also can fit the TRPL measured at F3 (Figure S11d). However, when the excitation fluence decreases to F2, the calculated TRPL decay becomes very fast without any slow component at all (because there is no trap saturation at this fluence). This behavior totally contradicts the experimentally observed kinetic which still has a very fast and very slow sections. These inconsistencies suggest that, under our experimental excitation conditions, the initial PL drop and the kinetics in general cannot be solely explained by saturation of deep traps.

## 6. Exponential tail of TRPL decay at long time scale at low excitation fluence: indication of first order processes at long timescale

In the presence of only shallow traps TRPL decay starts to follow the power law at long times. However, experimentally we observed that PL decay excited at $F_2$ fluence displays exponential decay at long timescale (Figure S12a). This is better visualized by plotting differential lifetime ($\tau_{diff} = -2 \cdot (\frac{d\, ln(I_{PL})}{dt})^{-1}$) as a function of time (Figure S12b).[7,8] This means that first order recombination processes become dominant at long time scales and should be considered. To account for this, we include the deep trap $T_D$ in the full model.

## 7. Indication of the presence of doping from a multi-pulse TRPL experiment

Here we consider PL response of a semiconductor to a burst of several short laser pulses.[5] The neighboring pulses are separated by equal time intervals (12.5 ns in our experiments), the bursts are repeated with very long repetition period (*e.g.* 1 ms).

In case of no permanent doping in the sample, and a very long repetition period of excitation bursts, the ratio between the initial PL intensity created by the first ($I_{PL1}$) and second ($I_{PL2}$) pulse can be easily estimated from the concentration $n_0$ of electrons and holes initially generated by one laser pulse:

$$I_{PL1} = k_r np = k_r \Delta n_0^2 \quad \text{...S7.1}$$

And for the second pulse

$$I_{PL2} = k_r(\Delta n_0 + n') \cdot (\Delta n_0 + p') \quad \text{(S7.2)}$$

Where $n'$ and $p'$ are the concentration of the remaining electrons and holes created by the first pulse at moment of the second pulse arrival.

We know that we have a strong photodoping effect in TC films and in perovskites in general that lasts at least hundreds of nanoseconds.[2–5] This is much longer than the distance between the pulses (12.5 ns in our case). That is why we can safely assume that the concentration of holes does not really change up to the moment of the second pulse arrival. At the same time, we can assume that all electrons have been trapped during the time until the arrival of the next laser pulse (it is the fast trapping of electrons that decreases the PL intensity by a factor of ≈ 6 times over 12.5 ns).

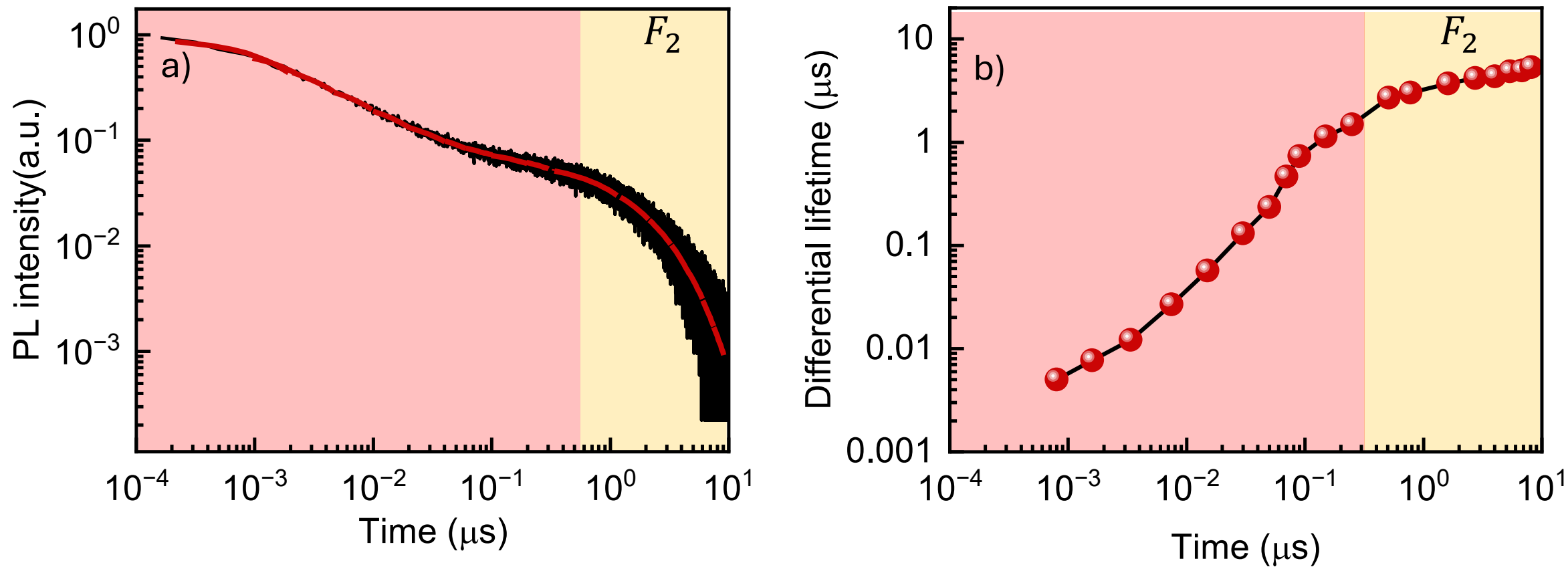

**Figure S12.** (a) Decay measured at low pulse fluence ($F_2$) shows exponential decay at long timescale, this is an indication of decay being dominated by first order processes there (yellow-marked time region), red line in this curve is piecewise fitted curve to a power law. (b) Differential lifetime as a function of time, it is almost constant in the yellow-marked time region.

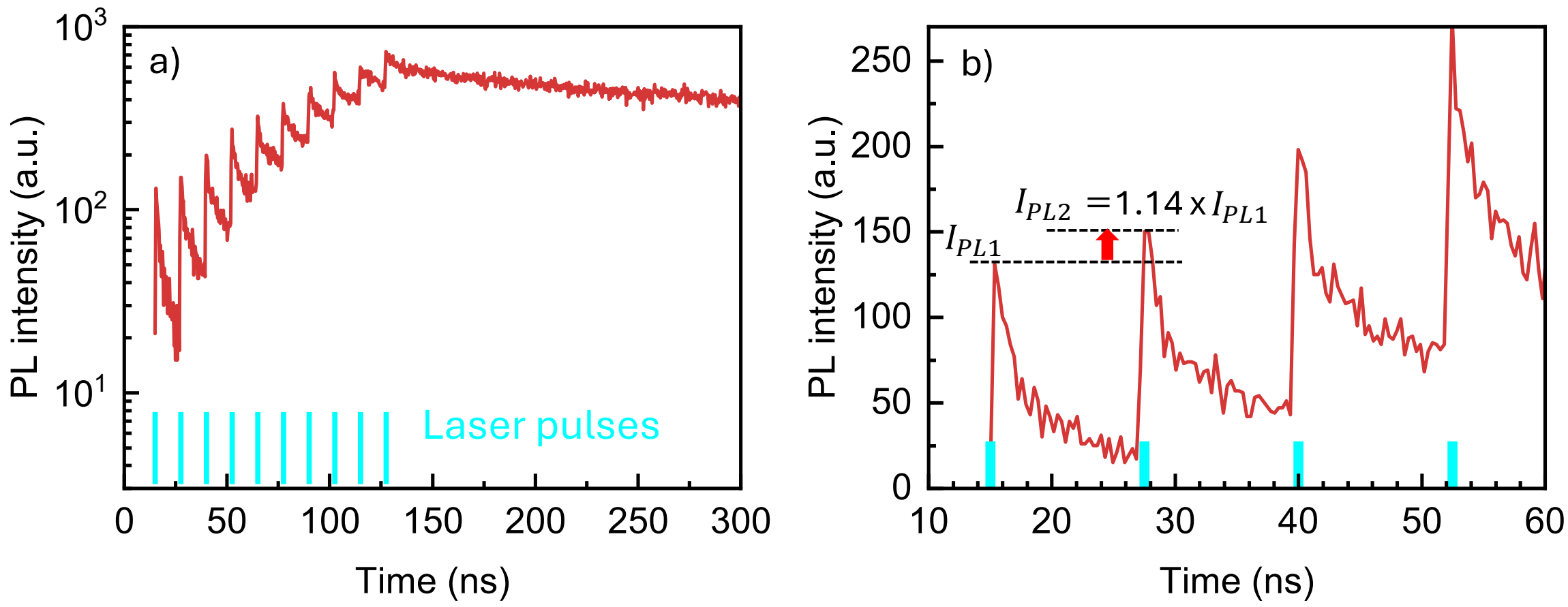

**Figure S13**. (a) TRPL decay under a muti-pulse excitation (pulse burst: 10 pulses with $F_2$ fluence at 80 MHz (12.5 ns between pulses), repetition period of the burst - 100 µs). (b) Zoom into the PL response to the first four pulses, the growth of the second PL response relative to the first one is shown.

In this case the initial PL intensity generated by the second pulse should be:

$$I_{PL2} = k_r \Delta n_0 (\Delta n_0 + \Delta n_0) = 2 I_{PL1} \quad \text{S7.3}$$

So, with the absence of chemical doping we expect the second PL response to be approximately 2 times larger than the first one. However, only 1.14 times increase was observed experimentally (Figure S13). This small increase can be readily explained assuming that there is always some concentration of holes present ($N_d$), in that case:

$$I_{PL1} = \Delta n_0 (\Delta n_0 + N_d)$$

$$I_{PL2} = \Delta n_0 (2\Delta n_0 + N_d)$$

By tuning $N_d$ any ratio $I_{PL2}/I_{PL1}$ in the range from 1 and 2 can be obtained.

## 8. Shockley–Read–Hall (SRH) model under low excitation power density

### Expected exponential TRPL decay under low excitation and single deep trap model

Under the condition of low excitation pulse fluence, equations of the SRH model can be written as:[3]

$$\frac{d}{dt} n(t) = -\, k_t\, N_t\, n \quad \text{(S8.1)}$$

$$\frac{d}{dt} n_{t1}(t) = \; k_t\, N_t\, n - k_n n_t p \quad \text{(S8.2)}$$

$$\frac{d}{dt} p(t) = \; -k_n n_t p \quad \text{(S8.3)}$$

With the initial conditions at the moment of the excitation laser pulse arrival:

$$n(0) = p(0) = \Delta n_0$$

$$n_t(0) = 0$$

Here, *n* and *p*, are the free electron and free hole densities in the conduction and valence bands, respectively. $n_0$ is the initial excitation density and $n_t$ is the electron density at the trap level. $k_t$, $N_t$, and $k_n$ are electron capture coefficient, trap density of electron traps, and hole capture coefficient (non-radiative recombination rate) respectively.
By solving equations S8.1, S8.2 and S8.3 we can find the electron and hole density in the conduction and valence band as a function of time, thereafter the PL intensity can be written in terms of $n$, and $p$ as,

$$I_{PL}(t) = k_r \cdot n(t) \cdot p(t) \tag{S8.4}$$

The solution of equation S8.1 can be written as

$$n(t) = \Delta n_0 \cdot \exp(-k_t N_t t) \tag{S8.5}$$

At low fluences the rate of trapping of electrons is much higher than the recombination rate of holes (because our trap is an electron trap with energy level shifted towards the conduction band, this makes capturing of an electron more probable than capturing a hole), therefore the whole decay that we see is due to disappearance of electrons, while the concentration of holes does not change. In this case the hole density can be written as,

$$p(t) = \Delta n_0 \tag{S8.6}$$

From Eq. S8.4, S8.5 and S8.6, we can find the expression of the PL intensity:

$$I_{PL}(t) = \ k_r np = k_r \Delta n_0^2 \cdot \exp(-k_t N_t t) \tag{S8.7}$$

Equation S8.7, shows that PL intensity decays according to mono-exponential law under low pulse fluence excitation.

## Exponential TRPL decay under low excitation in case of multiple deep traps

In the presence of many deep trap levels (index 1,2,3,…) the equations of the SRH model under low excitation condition can be written as

$$\frac{d}{dt} n(t) = -\ k_{t1}\, N_{t1}\, n - \ k_{t2}\, N_{t2}\, n - \ k_{t3}\, N_{t3}\, n \ldots \tag{S8.8}$$

$$\frac{d}{dt} n_{t1}(t) = \ -k_{t1}\, N_{t1}\, n - k_{n1} n_{t1} p \tag{S8.9}$$

$$\frac{d}{dt} n_{t2}(t) = \ -k_{t2} N_{t2}\, n - k_{n2} n_{t2} p \tag{S8.10}$$

$$\frac{d}{dt} n_{t3}(t) = \ k_{t2}\, N_{t3}\, n - k_{n3} n_{t3} p \tag{S8.11}$$

$$\vdots$$

$$\frac{d}{dt} p(t) = \ -k_{n1} n_{t1} p \ - k_{n2} n_{t1} p - k_{n3} n_{t1} p \tag{S8.12}$$

$$\frac{d}{dt} n(t) = -\ (k_{t1}\, N_{t1} + \ k_{t2}\, N_{t2} + \ k_{t3}\, N_{t3} + \cdots)\, n = \ -\textstyle\sum_i k_{ti} N_{ti}\, n \tag{S8.13}$$

$$\frac{d}{dt} p(t) = \ -(k_{n1} n_{t1} p + k_{n2} n_{t1} p + k_{n3} n_{t1} p + \cdots) p = \ -\textstyle\sum_i k_{ni} n_{ti}\, p \tag{S8.14}$$

the solution of Eq. S8.8 is:

$$n(t) = \Delta n_0 \cdot \exp(-\sum_i (k_{ti} N_{ti}) \cdot t) \quad \text{(S8.15)}$$

Following the same reasoning as in the previous section, we can assume that at low pulse fluence $p(t)$ is a constant $p(t) = \Delta n_0$
and the PL intensity can be written as

$$I_{PL}(t) = \Delta n_0^2 \cdot \exp(-\sum_i (k_{ti} N_{ti}) \cdot t) \quad \text{(S8.17)}$$

Equation S8.16 shows that in the presence of many deep trap levels the PL intensity will decay with the characteristic time $1/\sum_i (k_{ti} N_{ti})$, however the PL intensity decay is still mono-exponential.

## 9. Full SRH model ($T_D$,$T_{Sh}$,$T_A$,A,$N_d$) able to explain PLQY(*R, F*) map and TRPL decays together

To avoid complicated notations in the equation below, all the parameters related to the deep trap ($T_D$), shallow trap ($T_{Sh}$) and Auger trap ($T_A$) will be denoted with subscripts 1, 2 and 3 respectively. The differential equations of the full model can be written as follows:

$$\frac{dn(t)}{dt} = -k_{t1}(N_{t1} - n_{t1})n - k_{t2}(N_{t2} - n_{t2})n - k_{e3}(N_{t3} - n_{t3})np + k_{dt2} n_{t2} - k_r np - k_a np^2 \quad \text{(S9.1)}$$

$$\frac{dn_{t1}(t)}{dt} = k_{t1}(N_{t1} - n_{t1})n - k_{n1} n_{t1} p \quad \text{(S9.2)}$$

$$\frac{dn_{t2}(t)}{dt} = k_{t2}(N_{t2} - n_{t2})n - k_{n2} n_{t2} p - k_{dt2} n_{t2} \quad \text{(S9.3)}$$

$$\frac{dn_{t3}(t)}{dt} = k_{e3}(N_{t3} - n_{t3})np - k_{n3} n_{t3} p \quad \text{(S9.4)}$$

$$\frac{d}{dt} p(t) = -k_r np - k_{n1} n_{t1} p - k_{n2} n_{t2} p - k_{n3} n_{t3} p - k_a np^2 \quad \text{(S9.5)}$$

along with the condition of charge neutrality,

$$n(t) + n_{t1}(t) + n_{t2}(t) + n_{t3}(t) = p(t)$$

This system of differential equations can be solved numerically under the following initial conditions at the arrival moment (t=0) of the first laser excitation pulse:

$$n(0) = p(0) = \Delta n_0$$

$$n_{t1}(0) = n_{t2}(0) = n_{t2}(0) = 0$$

Where $N_{t1}$, $N_{t1}$, and $N_{t3}$ are the concentrations of the easily saturable deep traps ($T_D$), shallow traps ($T_{Sh}$) and Auger traps ($T_A$) respectively, all these traps are electron traps. $n_{t1}$ $n_{t2}$, and $n_{t3}$ are the densities of trapped electrons at $T_1$, $T_2$, and $T_3$ respectively. $k_{t1}$, $k_{t2}$ are the electron capture coefficients for $T_D$ and $T_{Sh}$, respectively. $k_{dt2}$ is de-trapping rate constant for electrons trapped at shallow traps, $k_{dt2}$ also can be expressed in terms of electron capture coefficient of the the shallow traps, effective density of state of the conduction band ($N_c$), energy level of the conduction band edge ($E_c$) and the trap energy level ($E_t$) as $k_{dt2} = k_{t2} N_c exp\left(-\frac{E_c - E_t}{k_t}\right)$. $k_{e3}$ is the Auger trapping rate constant of $T_A$ (process: n,p,empty trap), $k_r$ is radiative rate constant, $k_a$ is the Auger recombination rate constant (process: *n,p,p*). $k_{n1}$, $k_{n2}$, and $k_{n3}$ are hole capture coefficients (non-radiative recombination rate constants) of trapped electron with free hole for traps $T_D$, $T_{Sh}$ and $T_A$ respectively.
In the presence of chemical doping the condition of charge neutrality is modified to

$$n(t) + n_{t1}(t) + n_{t2}(t) + n_{t3}(t) + N_d = p(t)$$

here $N_d$ is the doping density, $N_d$ is negative in case of electron doping and positive in case of hole doping. In this case the initial conditions for solving the differential equations become

$$n(0) = \Delta n_0$$

$$p(0) = \Delta n_0 + N_d$$

$$n_{t1}(0) = n_{t2}(0) = n_{t3}(0) = 0$$

The PL intensity can be calculated by:

$$I_{PL}(t) = k_r n(t) p(t) \quad \text{(S9.6)}$$

Where $n(t), p(t)$ are obtained by solving equations from S9.1 – S9.5
To calculate PL response in the single pulse excitation regime we need to solve these equations only ones. This is because the response to the second pulse will be the same (no memory). However, this is only true for very low repetition rates.
If the system does not relax completely until the arrival of the second pulse, we need to solve this system of equations under modified initial conditions many times until a quasi-steady-state response is obtained (an equilibrium is reached). From this equilibrated level PLQY is then calculated. See more details elsewhere.[2–4]
The list of model parameters of the full model able to fit the full set of experimental data is given in Table S3.

## 10. Overview of simpler theoretical models explaining incomplete data sets

In this section we apply several simpler models (with lesser model parameters compared to the full model) to explain a part of the full PL data. The goal here is to find models which can explain a part of the PL data, while missing important physical processes and, thus, not able to explain the full data set. In practice, we have been considering all these models (and many other variants) until we came to the full model that fitted the data.

### 10.1 Model SRH-1: deep trap with normal trapping and Auger trapping, along with Auger recombination. Fitting only PLQY(*R, F*) map while ignoring low fluence PL decays

Steady state PL intensity and PLQY measurements only represent the result of time averaged charge carrier dynamics. That's why PLQY(*R, F*) can be explained even by ignoring some of the important processes responsible for peculiarities of the time-resolved PL data.

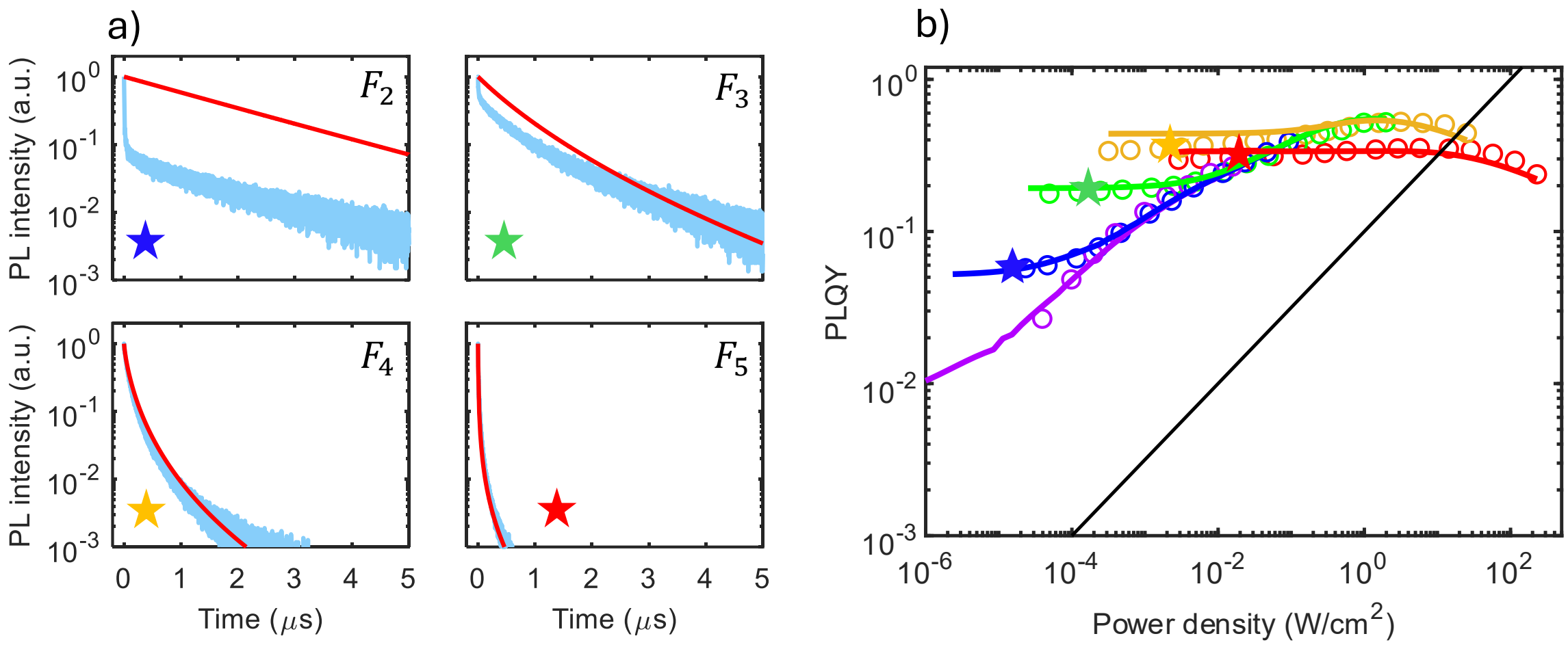

**Figure S14.** The figure shows the ability of the SRH model with only deep trap (model SRH-1) to nicely fit PLQY(*R, F*) map, and, at the same time, its failure to fit TRPL at low-middle fluences. This model was used to satisfactory fit the same type of data for $MAPbI_3$ films.[3]

For example, we can explain the full PLQY(*R, F*) map in the framework of a SRH model considering only one deep trap that can also trap electrons by the mechanism of Auger trapping, along with radiative and Auger recombination (total number of parameters is 6, see Table S6). This model has been successfully used previously to explain the same complex experimental data for $MAPbI_3$.[3] Note that the deep trap in this case is different than the deep trap and the Auger trap used in the full model, and we consider this case here for demonstration purposes only. Even though this model fits PLQY(*R, F*), it fails to display the initial fast component of the decay (as shown in Figure S14a). Note that this model can still explain the decay at high ($F_5$) and medium ($F_4$) fluences but starts to fail upon decreasing the fluence further as soon as the PL decay starts to show two (fast and slow) distinct components.

### 10.2 Model SRH-2: shallow trap only. Fitting low fluence TRPL decays with fast and slow components

The presence of slow and fast components in the TRPL decays can be explained by several physical processes, such as charge carrier diffusion, inhomogeneity of the sample, presence of shallow traps or saturation of the deep traps. As discussed in the main text, the presence of shallow traps is the most relevant to account for the slow and fast components in the TRPL decays of the TC perovskite films measured at low fluence ($F_2$).

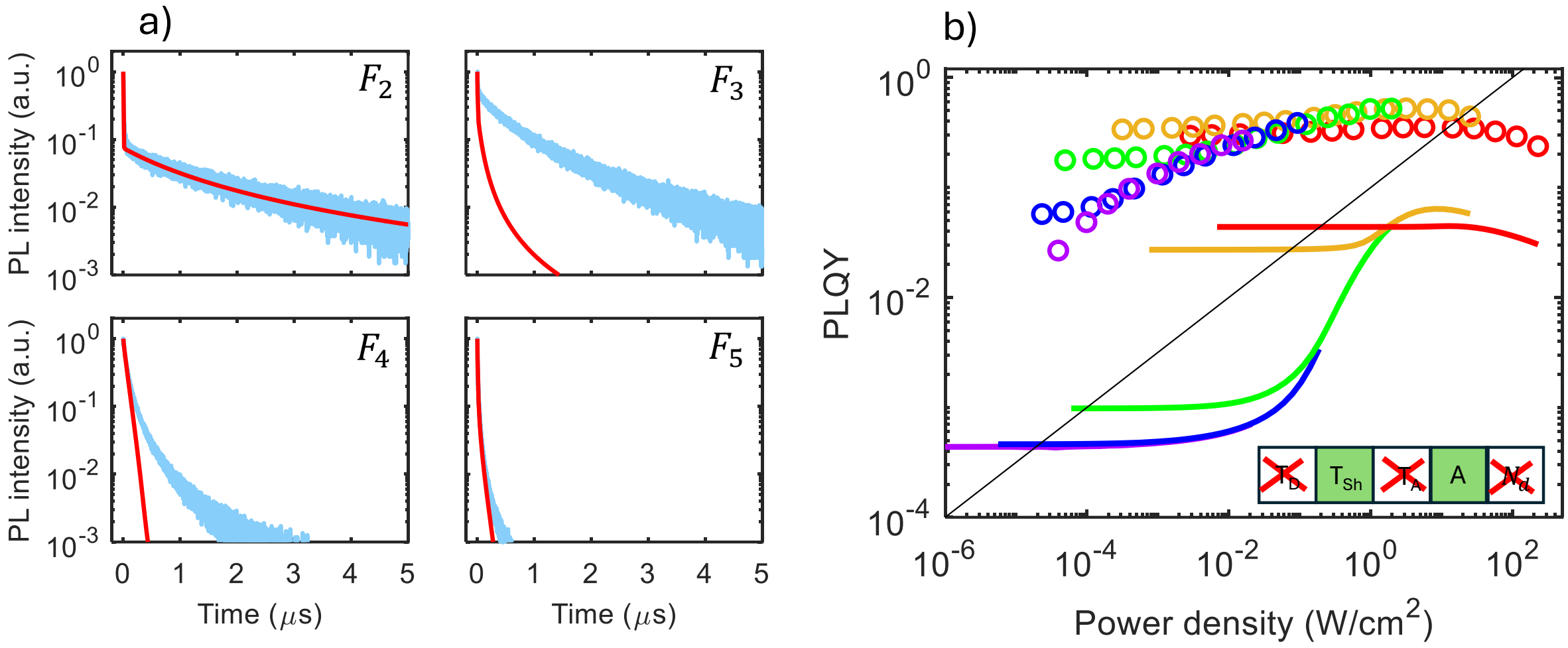

**Figure S15.** Fitting the PL decay at $F_2$ excitation fluence by the SRH-2 model, which assumes only one shallow trap (see Table S7 for the parameters). Note that all other data cannot be explained.

As shown in Figure S15a, the PL decay at low fluence $F_2$ can be fitted with just one shallow trap, however, this model (see Table S7 for the parameters) totally fails to predict all other features of the data.
The examples shown in Figure S14 and S15 highlight the importance of measuring and analyzing the PL decays over a broad range of excitation power density. The range of excitation power density must be broad enough so that the shape of the decays changes significantly in going from one extreme to another.

### 10.3 Model SRH-3: shallow trap, deep trap, Auger recombination. Fitting normalized TRPL decays while ignoring PLQY

In the sections 10.1 and 10.2 we highlighted the importance of measuring decays over a broad range of excitation fluence. However, even measuring TRPL decays over a very broad range of excitation fluence is not enough to retrieve the correct model if the PL decays are normalized (PLQY information is lost). Figure S16 shows that the PL decays over the whole fluence range can be approximately fitted by a SRH model considering one shallow and one deep trap and Auger recombination (see Table S8 for the parameters). However, this model does not fit the PLQY(*R, F*) map when forced to fit PL decays.

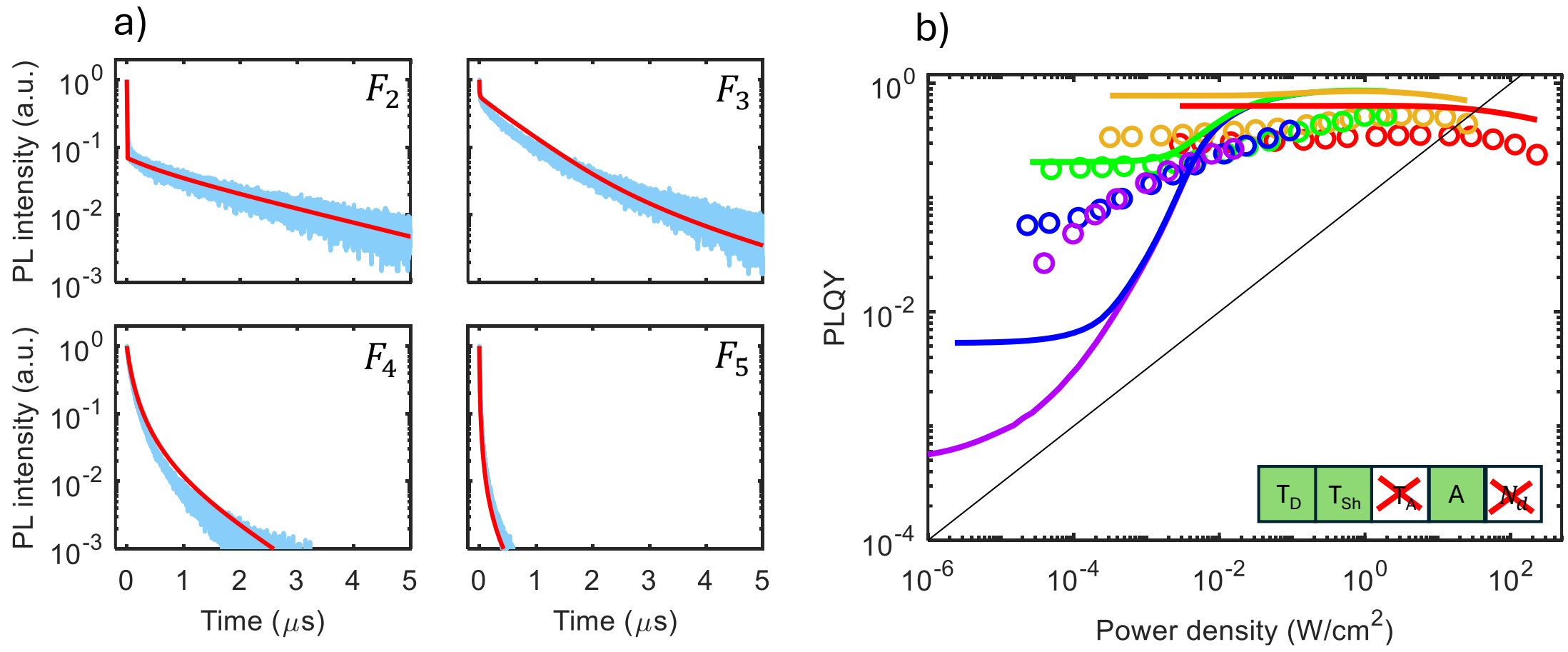

**Figure S16.** a) Experimental PL decays fitted by the SRH-3 model (one shallow, one deep trap and Auger recombination). b) The model with the same parameters fails, however, to predict the experimental PLQY(*R, F*) map. For the parameters see Table S8.

### 10.4 Model SRH-4: shallow trap, deep trap with normal trapping and Auger trapping, Auger recombination. Attempts to fit PLQY ($R$, $F$) and PL decays together

Here we use SRH-1 model with added shallow trapping. So, the model has one shallow and one deep trap, which can capture electrons directly and by Auger trapping mechanism. (Note that this deep trap combines normal and Auger trapping, which is different to the full model, where we considered 2 deep traps: first with only normal trapping the second with only Auger trapping). This model is complex enough to capture all features of the full data set (Figure 1, main text). Figure S17 shows a compromised fit of the TRPL decays and the PLQY map using the parameters given in Table S9. It qualitatively predicts all features of the data, fits the entire PLQY map and the PL decays at $F_5$ and $F_4$ fluences. As for the low fluence ($F_2$), it predicts the presence of an initial fast component which is later followed by a slow component. It also fits well the decay rate of the latter. However, it does not provide quantitative agreement with the amplitude of the initial drop of the PL intensity at $F_2$ fluence.

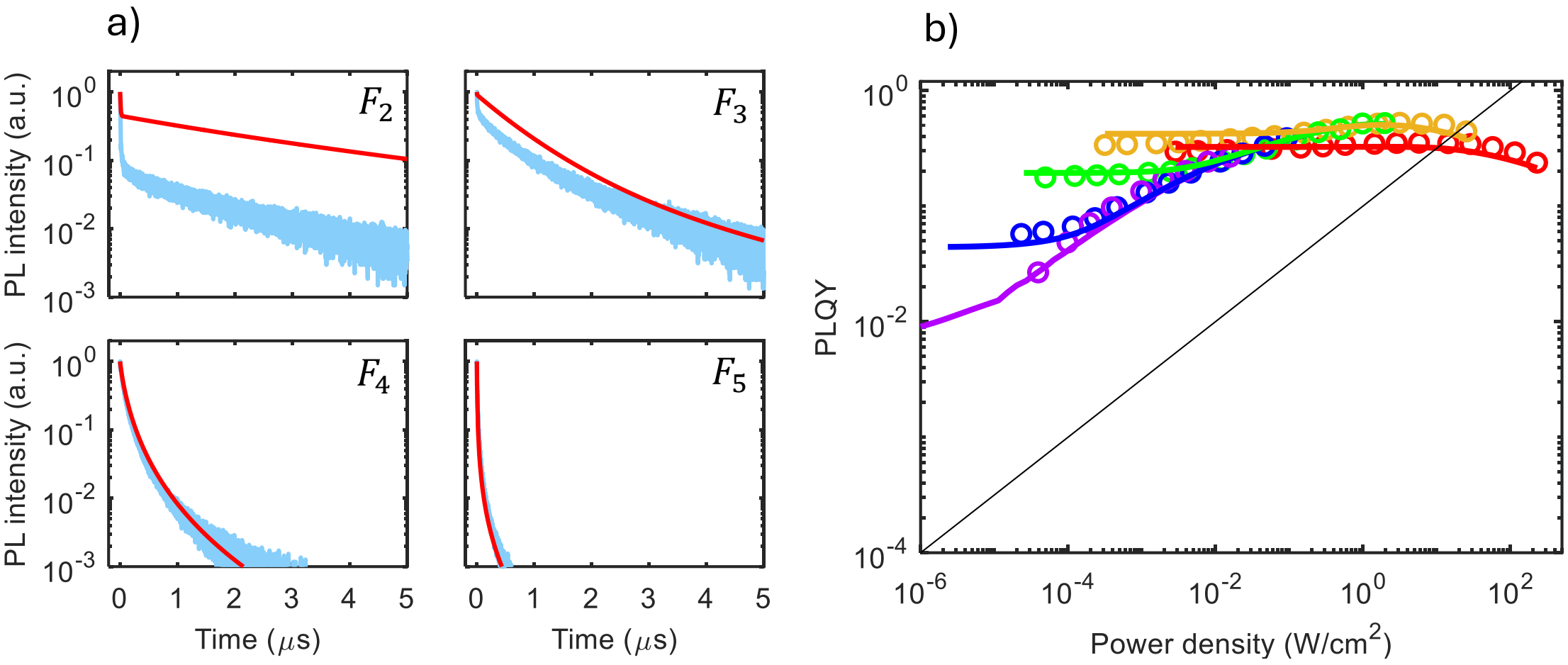

**Figure S17.** Fitting decays and PLQY($R$, $F$) map together using a SRH model with one shallow, and one deep trap (model SRH-4) which traps electrons directly and by Auger trapping mechanism. The model successfully predicts all the features of the data, failing, however, to predict the amplitude to the initial PL drop in the TRPL decay at pulse fluence $F_2$ and $F_3$.

We also found other set of parameters of the same model (Table S10), which allows fitting the amplitude of the initial drop. However, in this case we obtained instead the discrepancy in the decay rate of the slow component ($F_2$ and $F_3$ excitation fluences), see Figure S18.

The only principal effect which is still missing in this model is chemical doping. Adding chemical doping brings us to the full model (section 9) able to fit the whole set of data.

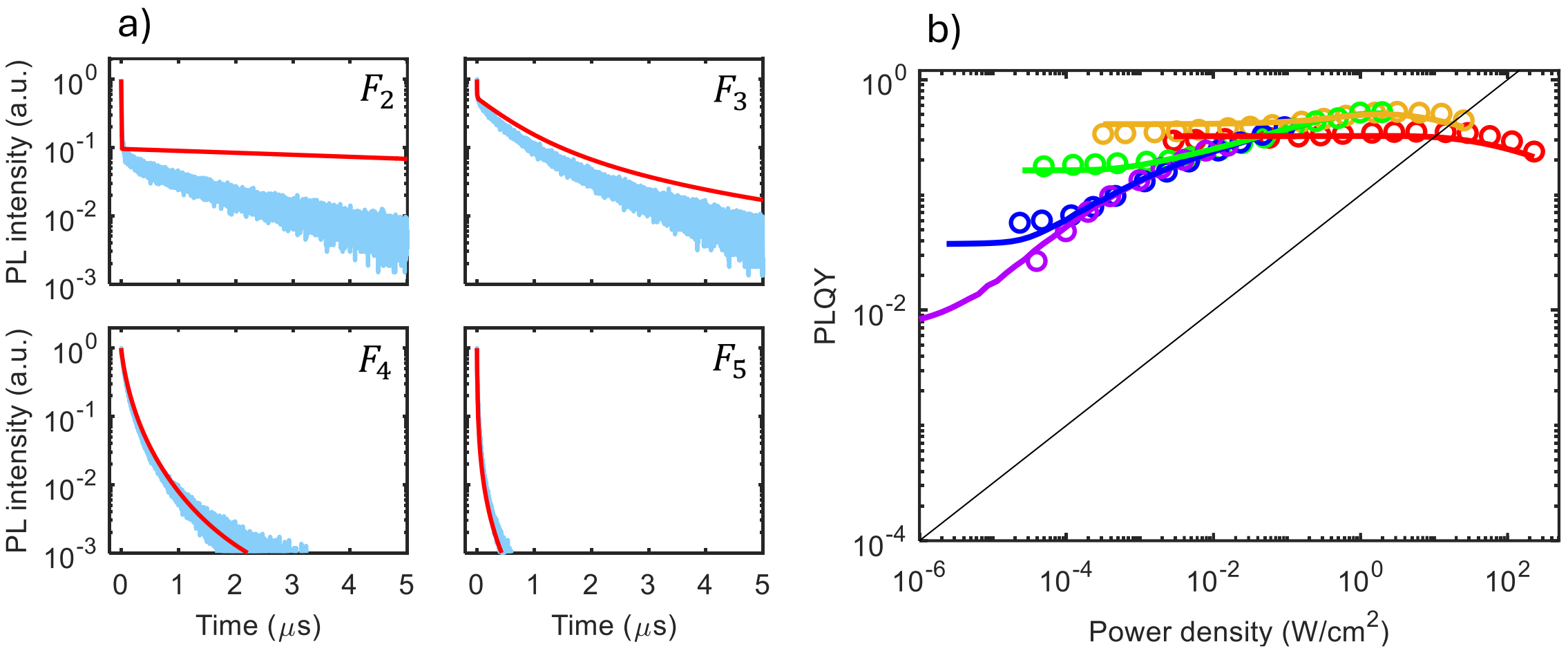

**Figure S18.** Fitting decays and PLQY(*R, F*) map together using the SRH-4 model, but with a different parameter set (Table S10) than in Figure S17. In this case it only fails to predict the decay rate of the long tail of TRPL measured at $F_2$.

## 11 List of model parameters

**Supplementary Table S3:** Table of model parameters of the full model used to explain decays and PLQY(*R, F*) altogether, (Figure 2 main text).

| Parameter | Value | Description |
|---|---|---|
| **Easily saturable deep trap ($T_D$)** | | |
| $k_{t1}N_{t1}$ ($1/k_{t1}N_{t1}$) | 1.00 x10$^{7}$ s$^{-1}$ (100 ns) | Electron capture rate constant of deep traps (SRH lifetime) |
| $N_{t1}$ | 2.50x10$^{14}$ cm$^{-3}$ | Concentration |
| $k_{n1}$ | 2.00x10$^{-11}$ cm$^{3}$s$^{-1}$ | Hole capture coefficient for deep traps |
| **Shallow trap ($T_{Sh}$)** | | |
| $k_{t2}N_{t2}$ ($1/k_{t2}N_{t2}$) | 4.80x10$^{8}$ s$^{-1}$ (2.08 ns) | Electron capture rate constant of shallow trap ($T_{Sh}$) (SRH lifetime) |
| $N_{t2}$ | 2.00x10$^{15}$ cm$^{-3}$ | Concentration |
| $k_{n2}$ | 1.60x10$^{-11}$ cm$^{3}$s$^{-1}$ | Hole capture coefficient for shallow traps |
| $k_{dt2}$ | 4.00x10$^{7}$ s$^{-1}$ | Electron de-trapping rate constant |
| **Auger trap ($T_A$), possesses Auger trapping only** | | |
| $N_{t3}$ | 8.00x10$^{15}$ cm$^{-3}$ | Concentration |
| $k_{e3}$ | 3.00x10$^{-26}$ cm$^{6}$s$^{-1}$ | Auger trapping rate constant |
| $k_{n3}$ | 1.60x10$^{-10}$ cm$^{3}$s$^{-1}$ | Hole capture coefficient for Auger traps |
| $k_r$ | 8.00x10$^{-11}$ cm$^{3}$s$^{-1}$ | Radiative recombination rate constant |
| $k_a$ | 8.00x10$^{-28}$ cm$^{6}$s$^{-1}$ | Auger recombination rate constant (*n,p,p* process) |
| $N_d$ | 3.27x10$^{15}$ cm$^{-3}$ | *p*-doping |

**Supplementary Table S4:** Table of model parameters for model SRH-0 (Figure S11a,b).

| Parameter | Value | Description |
|---|---|---|
| **Saturable deep trap** | | |
| $k_{t1}N_{t1}$ ($1/k_{t1}N_{t1}$) | 3.14x10$^{8}$ s$^{-1}$ (3.19 ns) | Electron capture rate constant of deep traps (SRH lifetime) |
| $N_{t1}$ | 3.20e+14 | Concentration |
| $k_{n1}$ | 3.27e-12 | Hole capture coefficient for deep traps |
| $k_r$ | 1.50e-09 | Radiative recombination rate constant |

**Supplementary Table S5:** Table of model parameters for model SRH-0 (Figure S11c,d).

| Parameter | Value | Description |
|---|---|---|
| **Saturable deep trap** | | |
| $k_{t1}N_{t1}$ ($1/k_{t1}N_{t1}$) | 1.57x10$^{8}$ s$^{-1}$ (6.37 ns) | Electron capture rate constant of deep traps (SRH lifetime) |
| $N_{t1}$ | 1.80e+15 | Concentration |
| $k_{n1}$ | 9.69e-13 | Hole capture coefficient for deep traps |
| $k_r$ | 3.80e-10 | Radiative recombination rate constant |

**Supplementary Table S6:** Table of model parameters for model SRH-1 (Figure S14).

| Parameter | Value | Description |
|---|---|---|
| **Deep trap possessing both direct trapping and Auger trapping** | | |
| $k_{t3}N_{t3}$ $1/k_{t3}N_{t3}$ | 4.00 x10$^{5}$ s$^{-1}$ 2.5 µs | Electron capture rate constant of deep traps (SRH lifetime) |
| $N_{t3}$ | 8.00x10$^{15}$ cm$^{-3}$ | Concentration |
| $k_{e3}$ | 3.00x10$^{-26}$ cm$^{6}$s$^{-1}$ | Auger trapping rate constant of deep traps |

| $k_{n3}$ | 1.60x10$^{-10}$ cm$^3$s$^{-1}$ | Hole capture coefficient |
|---|---|---|
| $k_r$ | 8.00x10$^{-11}$ cm$^3$s$^{-1}$ | Radiative recombination constant |
| $k_a$ | 8.00x10$^{-28}$ cm$^6$s$^{-1}$ | Auger recombination rate constant (*n*+*p*+*p* process) |

**Supplementary Table S7:** Table of parameters for model SRH-2 (Figure S15).

| Parameter | Value | Description |
|---|---|---|
| **Shallow trap** | | |
| $k_{t2}N_{t2}$<br>$(1/k_{t2}N_{t2})$ | 3.60x10$^8$ s$^{-1}$<br>(2.78 ns) | Electron capture rate constant of shallow trap<br>(SRH lifetime) |
| $N_{t2}$ | 4.00x10$^{15}$ cm$^{-3}$ | Concentration |
| $k_{n2}$ | 1.60x10$^{-9}$ cm$^3$s$^{-1}$ | Hole capture coefficient of shallow trap |
| $k_{dt2}$ | 2.50x10$^7$ s$^{-1}$ | Detrapping rate constant |
| $k_r$ | 1.00x10$^{-11}$ cm$^3$s$^{-1}$ | Radiative recombination rate constant |

**Supplementary Table S8:** Table of parameters for model SRH-3 (Figure S16).

| Parameter | Value | Description |
|---|---|---|
| **Deep trap** | | |
| $k_{t1}N_{t1}$<br>$(1/k_{t1}N_{t1})$ | 1.11x10$^7$ s$^{-1}$<br>(90.1 ns) | Electron capture rate constant of deep traps<br>(SRH lifetime) |
| $N_{t1}$ | 3.50x10$^{14}$ cm$^{-3}$ | Concentration |
| $k_{n1}$ | 7.80x10$^{-10}$ cm$^3$s$^{-1}$ | Hole capture coefficient of deep traps |
| **Shallow trap** | | |
| $k_{t2}N_{t2}$<br>$(1/k_{t2}N_{t2})$ | 6.00 x10$^8$ s$^{-1}$<br>(1.67 ns) | Electron capture rate constant of shallow trap<br>(SRH lifetime) |
| $N_{t2}$ | 1.50x10$^{15}$ cm$^{-3}$ | Concentration |
| $k_{n2}$ | 1.00x10$^{-13}$ cm$^3$s$^{-1}$ | Hole capture coefficient of shallow traps |
| $k_{dt2}$ | 3.40x10$^7$ s$^{-1}$ | Detrapping rate |
| $k_r$ | 1.40x10$^{-10}$ cm$^3$s$^{-1}$ | Radiative recombination rate constant |
| $k_a$ | 4.00x10$^{-28}$ cm$^6$s$^{-1}$ | Auger recombination rate constant |

**Supplementary Table S9:** Model SRH-4, table of parameters used in Figure S17.

| Parameter | Value | Description |
|---|---|---|
| **Shallow trap** | | |
| $k_{t2}N_{t2}$ <br> $(1/k_{t2}N_{t2})$ | 9.60 x$10^{7}$ $s^{-1}$ <br> (10.4 ns) | Electron trapping rate constant of shallow trap <br> (SRH lifetime) |
| $N_{t2}$ | 4.00x$10^{14}$ $cm^{-3}$ | Concentration |
| $k_{n2}$ | 9.60x$10^{-11}$ $cm^{3}s^{-1}$ | Hole capture coefficient of shallow traps |
| $k_{dt}$ | 4.00x$10^{7}$ $s^{-1}$ | Detrapping rate |
| **Deep trap possessing both direct trapping and Auger trapping** | | |
| $k_{t3}N_{3}$ <br> $(1/k_{t3}N_{t3})$ | 4.00x$10^{5}$ $s^{-1}$ <br> 2.5 µs | Electron trapping rate constant of deep traps <br> (SRH lifetime) |
| $N_{t3}$ | 8.00x$10^{15}$ $cm^{-3}$ | Concentration |
| $k_{e3}$ | 2.00x$10^{-26}$ $cm^{6}s^{-1}$ | Auger trapping rate constant |
| $k_{n3}$ | 2.40x$10^{-10}$ $cm^{3}s^{-1}$ | Hole capture coefficient of deep traps |
| $k_{r}$ | 8.00x$10^{-11}$ $cm^{3}s^{-1}$ | Radiative recombination rate constant |
| $k_{a}$ | 8.00x$10^{-28}$ $cm^{6}s^{-1}$ | Auger recombination rate constant |

**Supplementary Table S10:** Model SRH-4, table of parameters used in Figure S18.

| Parameter | Value | Description |
|---|---|---|
| **Shallow trap** | | |
| $k_{t2}N_{t2}$ <br> $1/k_{t2}N_{t2}$ | 4.80 x$10^{8}$ $s^{-1}$ <br> (2.08 ns) | Electron trapping rate constant of shallow trap <br> (SRH lifetime) |
| $N_{t2}$ | 2.00x$10^{15}$ $cm^{-3}$ | Concentration |
| $k_{n2}$ | 1.60x$10^{-11}$ $cm^{3}s^{-1}$ | Hole capture coefficient of shallow traps |
| $k_{dt}$ | 4.00x$10^{7}$ $s^{-1}$ | Detrapping rate |
| **Deep trap possessing both direct trapping and Auger trapping** | | |
| $k_{t3}N_{t3}$ <br> $1/k_{t3}N_{t3}$ | 4.00x$10^{5}$ $s^{-1}$ <br> (2.5 µs) | Electron trapping rate constant of deep traps <br> (SRH lifetime) |
| $N_{t3}$ | 8.00x$10^{15}$ $cm^{-3}$ | Concentration |
| $k_{e}$ | 2.00x$10^{-26}$ $cm^{6}s^{-1}$ | Auger trapping rate constant |
| $k_{n3}$ | 2.40x$10^{-10}$ $cm^{3}s^{-1}$ | Hole capture coefficient of deep traps |
| $k_{r}$ | 8.00x$10^{-11}$ $cm^{3}s^{-1}$ | Radiative recombination rate constant |
| $k_{a}$ | 8.00x$10^{-28}$ $cm^{6}s^{-1}$ | Auger recombination rate constant |